\documentclass[twocolumn,10pt]{article}
\usepackage{cite}
\usepackage[cmex10]{amsmath}
\usepackage{url}

\usepackage{units}
\usepackage[super]{nth}
\usepackage{multirow}
\usepackage{algorithm}
\usepackage[noend]{algpseudocode}
\usepackage{graphicx}
\graphicspath{ {.} }
\usepackage{xcolor}
\usepackage{comment}
\usepackage{balance}
\usepackage[]{subcaption}
\usepackage{orcidlink}

\begin{document}
\title{Subverting Stateful Firewalls with Protocol States\\
(Extended Version)\footnote{This is an extended version of a paper that will be published in NDSS 2022.}
}

\author{Amit Klein
\orcidlink{0000-0002-8024-8756}\\
Bar Ilan University\\
aksecurity@gmail.com
}

\date{} 
\maketitle

\begin{abstract}
We analyzed the generation of protocol header fields in the implementations of multiple TCP/IP network stacks and found new ways to leak information about global protocol states. We then demonstrated new covert channels by {\em remotely} observing and modifying the system's global state via these protocol fields. Unlike earlier works, our research focuses on hosts that reside in {\em firewalled} networks (including source address validation -- SAV), which is a very common scenario nowadays. Our attacks are designed to be {\em non-disruptive} -- in the exfiltration scenario, this makes the attacks stealthier and thus extends their longevity, and in case of host alias resolution and similar techniques -- this ensures the techniques are ethical. We focused on ICMP, which is commonly served by firewalls, and on UDP, which is forecasted to take a more prominent share of the Internet traffic with the advent of HTTP/3 and QUIC, though we report results for TCP as well.

The information leakage scenarios we discovered enable the construction of practical covert channels which directly pierce firewalls, or indirectly establish communication via hosts in firewalled networks that also employ SAV. We describe and test three novel attacks in this context: exfiltration via the firewall itself, exfiltration via a DMZ host, and exfiltration via co-resident containers. These are three generic, new use cases for covert channels that work around firewalling and enable devices that are not allowed direct communication with the Internet, to still exfiltrate data out of the network. In other words, we exfiltrate data from isolated networks to the Internet. We also explain how to mount known attacks such as host alias resolution, de-NATting and container co-residence detection, using the new information leakage techniques.
\end{abstract}

\newcommand{\IP}{\mathit{IP}}
\newcommand{\ID}{\mathit{ID}}
\newcommand{\src}{\mathit{SRC}}
\newcommand{\dst}{\mathit{DST}}
\newcommand{\target}{\mathit{TARGET}}
\newcommand{\receiver}{\mathit{RCVR}}
\newcommand{\UDP}{\mathit{IPPROTO\_UDP}}
\newcommand{\RTO}{\mathit{IRTO}}

\section{Introduction}

Our research focuses on how to exploit global state in the UDP and ICMP protocol stacks to mount attacks against firewalled networks and hosts. The attacks we are interested in are data exfiltration, host alias resolution and de-NATting, and co-residency detection -- see Table~\ref{tab:use-cases}. In some cases, the same attacks can be carried out over TCP, in a more efficient manner than existing techniques.

\subsection{Motivation}
Our motivating example is a highly secured network, wherein internal machines are not allowed to communicate directly with the Internet. This is enforced by a network firewall, which prevents the network hosts from accessing the Internet, but may allow them to access services on a neighboring internal network and/or management access to the firewall itself. Such {\em isolated} networks can be found in financial institutions, defense industries, hospitals, sensitive infrastructure sites, etc. The attacker's goal is to exfiltrate data out of an isolated network, to the Internet. The compromised data may consist of small ``strategic'' pieces of data, such as cryptographic keys, or compressed text files. 

We assume that an attacker controls a compromised internal machine. For example, the attacker may have installed malware on that machine, by means out of scope for this research. 
The attacker's goal is to exfiltrate data from this machine. It should be noted that the machine may be part of an attacker network of compromised machines, such that the attacker funnels the data to be exfiltrated from other machines to the machine from which the exfitration is to take place.

Our techniques are able to circumvent the firewall's enforcement, and allow the attacker to exfiltrate data from a machine in an isolated network to an Internet host the attacker controls. The exfiltration throughput is hundreds to thousands bits per hour (except for Windows), which suffices to transmit e.g. a 128-bit key in less than 22 minutes.

It should be stressed that this is an example to one use case (firewall circumvention); in the paper we also outline other attacks, such as host alias resolution and de-NATting, and co-residency detection. The motivation for these use cases is to provide operational intelligence as part of a larger attack effort. Detecting that two services reside on the same host may enable an attack against a weaker (vulnerable) service to take over a host which also runs a more strategic (but not vulnerable in itself) service.
\begin{table}
\caption{Attack Use Cases}
\begin{center}
\resizebox{.5\textwidth}{!}{
\begin{tabular}{|l|l|l|l|}
\hline
Use Case                                                                       & \begin{tabular}[c]{@{}l@{}}Sender\\Location\end{tabular}                                                     & \begin{tabular}[c]{@{}l@{}}Target Host\end{tabular}                                      & \begin{tabular}[c]{@{}l@{}}Receiver\\Location\end{tabular}                                                     \\ \hline
\begin{tabular}[c]{@{}l@{}}Firewall Piercing\\ (exfiltration)\end{tabular}           & \begin{tabular}[c]{@{}l@{}}Isolated\\ network\end{tabular} & Firewall                                                                                   & Internet                                                     \\ \hline
\begin{tabular}[c]{@{}l@{}}Exfiltration\\(via DMZ)\end{tabular}                                                                   & \begin{tabular}[c]{@{}l@{}}Isolated\\ network\end{tabular} & \begin{tabular}[c]{@{}l@{}}Firewalled\\ host in DMZ\end{tabular}                           & Internet                                                     \\ \hline
\begin{tabular}[c]{@{}l@{}}Co-Resident\\ Container\\ Exfiltration\end{tabular} & \begin{tabular}[c]{@{}l@{}}Isolated\\ network\end{tabular} & \begin{tabular}[c]{@{}l@{}}Two co-resident\\ containers\\ (internal+external)\end{tabular} & Internet                                                     \\ \hline
\begin{tabular}[c]{@{}l@{}}Alias Detection\\ and de-NATting\end{tabular}       & Internet                                                   & \begin{tabular}[c]{@{}l@{}}Firewalled/NATted\\ host\end{tabular}                           & \begin{tabular}[c]{@{}l@{}}Internet\\ (=Sender)\end{tabular} \\ \hline
\end{tabular}
}
\end{center}
\label{tab:use-cases}
\end{table}

\subsection{Organization}
Next in this section we provide background on terms and concepts we use throughout the paper (stateful firewalls, host alias resolution, and co-residency), followed by a summary of our contributions. In Section~\ref{related-work} we discuss related work. Section~\ref{sec:current-protocols} describes the current algorithms that generate the protocol fields of interest. Section~\ref{sec:attacks} outlines our attack concepts, Section~\ref{sec:channel-details} then explains how the concepts are implemented per-OS, and Section~\ref{sec:experiments} describes our experiments. A discussion of remediation options is provided in Section~\ref{sec:remediation}. We draw our conclusions in Section~\ref{sec:conclusions}, and finally we report on the vendor status in Section~\ref{sec:vendor-status}.

\subsection{Stateful Firewalls and SAV}
A {\em stateful firewall} is a ``network-based firewall that individually tracks sessions of network connections traversing it'' \cite{wiki:stateful-firewall}. According to \cite{10.5555/1207185}, ``[a] stateful inspection firewall is the de facto standard for network protection''. SAV, and some features of stateful firewalls can eliminate known data leakage attacks. We explain this below, and we note that our techniques are not affected by SAV and by stateful firewalls.

In response to an incoming invalid packet -- a TCP packet or a UDP packet destined to a closed port, or a TCP packet (other than SYN-only to an open port) that does not belong to an already established TCP session -- a firewall may either silently {\em drop} the packet, or actively {\em reject} it. Rejection is carried out by sending back a TCP RST packet in the TCP case, or by sending back an ICMP Type 3 Code 3 packet (``port unreachable'') in the UDP case.

Dropping invalid packets is considered a somewhat more secure choice (at the expense of a slight overall performance degradation due to a delay in responsiveness at the remote party when no rejection packet is sent), and is oftentimes set as the default firewall behavior. 
In Appendix~\ref{app:probing}, we report the results of an experiment in which we measured and estimated the portion of widely-used Internet hosts that do not send any response for UDP and TCP packets arriving to closed ports. We found that the vast majority of the Internet hosts silently drop such UDP/TCP packets. Therefore, attacks that assume a response for an invalid, connecion-less UDP/TCP packet will fail for most servers. 
Our attacks, on the other hand, only use valid and acceptable packets, and are thus unaffected by stateful firewalls.

A firewall can also employ Source Address Validation (SAV), such that it does not allow traffic from a first network into a second network with source addresses outside the first network. This security measure can prevent some trivial information leaking attacks. For example, a host $A$ on an isolated, internal network which is prevented from sending packets to the Internet can send a spoofed packet whose source address $C$ is an Internet address, to host $B$ in the DMZ network which is allowed to send data to the Internet. Host $B$ will then respond by sending a packet to $C$, an Internet address. This can be used as a covert channel signal. SAV prevents this by dropping $A$'s spoofed packets since their spoofed source address $C$ does not match their origin (the internal network).

Henceforth, we use the term {\em firewall} as a shorthand for ``a stateful firewall that silently drops invalid packets, with SAV between all its networks''. 

\subsection{Host Alias Resolution, De-NATting and Co-Residency Detection}
Note:
    we use the term {\em host} throughout this paper to denote a single instance of an operating system kernel. This can be a physical host, or -- if virtual machines are used -- a single virtual machine. We use the term {\em physical host} to explicitly exclude the latter case.
    
A {\em host alias resolution} \cite{10.1109/TNET.2003.822655} (also {\em host alias detection} \cite{10.1007/978-3-540-31966-5_9}) technique enables a remote party to infer whether two given IP addresses map to the same host or to two different hosts. We can extend this into inferring whether two services (defined by IP address, the listening port and the transport protocol -- TCP or UDP) map to the same host or not. For example, one can ask whether 1.2.3.4:80 (over TCP) and 5.6.7.8:443 (over UDP) map to the same host.

{\em De-NATting} \cite{Bellovin2002} \cite{Herzberg2017} is a use case of host alias detection, wherein the host(s) in questions reside behind a NAT, and thus the two services are mapped from their external endpoints into internal endpoints by a NAT device. 

In this research, we focus on {\em server} de-NATting, meaning, discerning whether two external endpoints refer to the same host or to two different hosts. This is in contrast to {\em client} de-NATting, which refers to discerning whether two outbound connections (designated by their external address endpoints) are made by a single client or by two different clients. The material difference is that a NATted client is likely to use a single internal IP address, whereas a NATted server may use two IP addresses for two different services. Some operating systems use the $(\IP_\src,\IP_\dst)$ tuple to generate network fields, making it trivial to de-NAT clients, but not servers.

{\em Co-Residency Detection for Containers} \cite{10.1145/3411495.3421357} is a concept similar to host alias detection, which aims to detect whether two endpoints on two different containers are hosted by the same physical host. If virtualization is used, then the containers may run on 
two different virtual machines. We are interested in a variant of this concept, which is to detect whether two endpoints on two different containers are hosted by the same {\em operating system kernel}.

Container technology (e.g. Docker) is available for Linux and Windows. Our co-residency detection attacks apply to Linux, but not to Windows.

\subsection{Our Contribution}

\begin{table}
\caption{Information Leaking Fields and Their Channel Properties}
\label{tab:main-results}
\setlength{\tabcolsep}{5pt}
\begin{center}
\resizebox{.5\textwidth}{!}{
\begin{tabular}{|l|l|l|l|l|}
\hline
OS      & Protocol                                                                   & Field                                                     & \begin{tabular}[c]{@{}l@{}}Sender\\ Capabilities\end{tabular} & Channel Bandwidth                                                                                                \\ \hline
Linux   & \begin{tabular}[c]{@{}l@{}}UDP/IPv4, \\ ICMP/IPv4\end{tabular}             & IPv4 ID                                                   & IP Spoofing            & \begin{tabular}[c]{@{}l@{}}400 b/h (UDP), \\ 3000 b/h (ICMP)\end{tabular}                                        \\ \hline
Windows & \begin{tabular}[c]{@{}l@{}}UDP/IPv4, \\ TCP/IPv4\end{tabular}              & IPv4 ID                                                   & IP Spoofing            & \begin{tabular}[c]{@{}l@{}}47.4 b/h (UDP), \\ 41.4 b/h (TCP)\end{tabular}                                        \\ \hline
macOS   & UDP/IPv4             & IPv4 ID                                                   & \begin{tabular}[c]{@{}l@{}}Web Traffic\\Emission\end{tabular}         & 1800 b/h                            \\ \hline
macOS   & ICMP/IPv4 & (rate limit)                                                   & \begin{tabular}[c]{@{}l@{}}Web Traffic\\Emission\end{tabular}         & 1800 b/h (estimated)                            \\ \hline
OpenBSD & \begin{tabular}[c]{@{}l@{}}UDP/IPv4, \\ TCP/IPv4 \\ ICMP/IPv4\end{tabular} & IPv4 ID                                                   & \begin{tabular}[c]{@{}l@{}}Web Traffic\\Emission\end{tabular}         & \begin{tabular}[c]{@{}l@{}}720 b/h (UDP, throttled), \\ 360 b/h (TCP, throttled) \\ 4200 b/h (ICMP)\end{tabular} \\ \hline
NetBSD  & \begin{tabular}[c]{@{}l@{}}TCP/IPv4, \\ TCP/IPv6\end{tabular}              & TCP ISN                                                   & \begin{tabular}[c]{@{}l@{}}Web Traffic\\Emission\end{tabular}         & 7200 b/h (throttled)                                                                                             \\ \hline
NetBSD  & TCP/IPv6                                                                   & \begin{tabular}[c]{@{}l@{}}IPv6 Flow\\ Label\end{tabular} & \begin{tabular}[c]{@{}l@{}}Web Traffic\\Emission\end{tabular}         & Untested                                                                                                         \\ \hline
\end{tabular}
}
\end{center}
\end{table}

\begin{itemize}
    \item New \textbf{information leakage techniques} based on the statefulness of IPv4 ID, TCP ISN (IPv4 and IPv6) and IPv6 flow label in popular server operating systems: Linux, Windows Server, macOS Server, OpenBSD and NetBSD. In the IPv4 ID and TCP ISN cases, this is done without accurately predicting the full values of these protocol fields. Our results are summarized in Table~\ref{tab:main-results}. These techniques are non-disruptive, and can be used even when the target host is firewalled, including SAV. The attacks are algorithmic, and therefore are not hardware-specific, nor do they require fine-grained timing or rely on race conditions. And since they rely on kernel implementations, they are agnostic to the nature of applications/services the operating system runs.
    \item Exfiltration from a firewalled network (including SAV) and exfiltration between co-resident containers  -- a new \textbf{covert channel use case} for exploitation of stateful IPv4 ID, TCP ISN and IPv6 flow label. In this use case, an internal device is prevented from directly connecting to the Internet, by a firewall, but is allowed to connect to the firewall itself, or to a DMZ host, or to an internal container co-residing with a \nth{3} party container. Our attacks exploit the host's global protocol states to exfiltrate data from the internal device to the Internet. The covert channels' bandwidth (in bits/hour) depends on the information leakage techniques' bandwidth listed in Table~\ref{tab:main-results}.
    \item Constructing a covert channel based on the new information leakage techniques and exploiting it for well known attack use cases: server alias resolution, de-NATting, container co-residency detection, and idle scanning. Previously, these attacks were inapplicable either due to the underlying information leaking technique being mitigated, or due to the disruptive nature of the underlying technique or other limitations.
        \item A description of the present day algorithm used by Windows to manage $(\IP_{\src},\IP_{\dst})$ tuples, including IPv4 ID management.
\end{itemize}

\section{Related Work}
\label{related-work}
\subsection{IPv4 Global State and Side Channels}
\subsubsection{Prediction of IPv4 ID values}
The IPv4 ID field is a 16 bit identifier used to reassemble IP packets from smaller fragments. This field is populated by the OS regardless of whether fragmentation is actually needed. If the IPv4 ID field is generated in a predictable manner, e.g. sequential increment per packet, then it is possible to sample the field twice and determine whether the target host sent another packet in between. For this approach, the IPv4 ID field must be highly predictable, i.e with a near-zero entropy. Note of course that if the ID is static (e.g. 0), which is legitimate for atomic (DF=1)\footnote{The IPv4 DF (``Don't Fragment'') header flag controls fragmentation by routers. Setting DF=1 prohibits routers from fragmenting the packet.} packets \cite[Section 4]{rfc6864}, then no information is leaked. However, some operating systems (e.g. Linux) do set DF=1 for short packets, yet still populate the ID field with a non-static value.

The original idle-scan technique \cite{idlescan} exploited sequential IP IDs which were very common in the operating systems of that era (1998). Later in 2008, \cite{BSD-IPID-attack} cryptanalyzed the IPv4 ID mechanism of OpenBSD and macOS Server and showed it to be predictable. In response, OpenBSD and macOS Server changed their algorithm. In 2019, \cite{KP19-usenix} attacked the IP ID generation in Windows TCP/IPv4 and UDP/IPv4, and in Linux UDP/IPv4. In response, Windows and Linux changed their algorithms. Recently, \cite{Portland} showed that the Linux TCP/IPv4 ID field is also predictable, and in response Linux changed that algorithm as well.

The only operating systems nowadays that generate predictable (sequential) IPv4 IDs are FreeBSD (UDP/IPv4 only) and NetBSD (only for packets longer than 68 bytes, thus excluding e.g. TCP RST packets).
With the exception of FreeBSD and NetBSD, all operating systems we surveyed upgraded their IPv4 ID generation scheme to a non-predictable algorithm, hence attacks that rely on IPv4 ID prediction are no longer applicable to them. 

Moreover, in general, idle-scanning, which relies on the pivot server to send an RST packet for ``unexpected'' SYN+ACK packet, does not apply to a firewalled pivot server. A firewall is likely to silently drop the SYN+ACK packet, or alternatively the firewall may respond with an RST packet, without forwarding it to the pivot server. We do, however, note that our NetBSD flow label attack can be used for {\em efficient} idle scanning TCP/IPv6 hosts with NetBSD as a pivot server, when the latter is not firewalled.

In 2008, Danezis described how to construct covert channels from incremental IPv4 IDs \cite{DBLP:conf/spw/Danezis08}. But Danezis's use case is different than ours. Danzeis assumes the monitored party can still connect to the Internet, and as such, he focuses on channel robustness, rather than on the technicalities of implementing the technique for firewalled networks. Additionally, that work does not describe UDP-based techniques, IPv6-based techniques and container-based techniques.

In contrast, the global IPv4 ID states we describe exist in modern IPv4 ID generator implementations. These global states are typically subtle and not easily observed.

\subsubsection{Attacks on the Linux IPv4 ID Generation Algorithm for UDP and TCP Connection-Less Packets}
The Linux IPv4 ID generation algorithm for {\em connection-less packets} (ICMP and UDP; until v4.19 also TCP RST packets not in a connection context) is described in Section~\ref{sec:linux-ipid-algo}. This construction was attacked in several recent papers. 
Some attacks against Linux TCP connections exploited this algorithm, either when used for TCP RST packets \cite{conf/infocom/ZhangKC18} \cite{DetectingTCPIPConnectionsviaIPIDHashCollisions} (note that since v4.19, the Linux kernel sets IPv4 ID to 0 in RST packets which are not in an existing connection context), or through a ``downgrade attack'' from the more secure TCP IP ID generation scheme \cite{feng2020off}. However, these attacks do not affect Linux UDP and ICMP traffic, which is the object of our research. Additionally, these attacks practically require the attacker to own thousands of IPv4 addresses. 
In 2019, \cite{KP19-usenix} exploited the Linux IP ID generation structure to find partial hash collisions and extract the hashing key. In response, the hash function was modified to use a larger key.

A recent DNS cache poisoning attack \cite{255314} involving Linux-based recursive resolvers hinges on the fact that when two machines are behind the same NAT, their external IPv4 address is identical, and hence the packets sent by a Linux host to both machines share the IPv4 ID counter.
But to exploit this for exfiltration, the sender and receiver must be on the same network and behind the same NAT, which is inapplicable to an exfiltration scenario.

Similarly to the above attacks, our attacks exploit the counter array of the Linux IPv4 ID algorithm, but in contrast to the previous attacks, our attacks target ICMP and UDP services, and they do not require ownership of thousands of IPv4 addresses, or having the sender and the receiver reside on the same network, and can target any host.

\subsubsection{IPv4 De-fragmentation Cache}
Another global IPv4 state is the memory limit for the de-fragmentation cache, a kernel data structure that maintains the incoming IPv4 fragments. The limit can be applied per source IP address, or globally. Overflowing the de-fragmentation cache can be detected remotely, as explained in \cite{Gilad:2013:FCV:2445566.2445568}, and can thus be used as a side channel. This technique requires IPv4 fragments to traverse the Internet unmodified. However, the prevalence of Carrier Grade NATs (CGNATs) reduces the attack surface of this technique. In CGNATs, 
the attacker cannot space the arrival time to the target host of fragments that belong to the same IP packet. Rather, the first fragment will await virtual de-fragmentation on the CGNAT device, and will only be transmitted when the second fragment arrives at the CGNAT. In fact, anecdotal evidence from our lab's ISP indicates that CGNATs may employ an even more destructive strategy, wherein they reassemble the IPv4 packet and forward the complete packet.

Additionally, since this attack overflows the de-fragmentation cache, it is disruptive, as organic inbound fragments may be evicted from the cache before re-assembly, resulting in dropping of organic inbound packets.

\subsection{UDP Global State and Side Channels}
The UDP protocol does not contain global states per-se. However, \cite{UDP-ICMP-scan} describes a global state based on a rate limit imposed by some kernels on ICMP protocol ``port unreachable'' messages, sent in response to incoming UDP packets destined to closed ports.
It should be noted though, that firewalls handle UDP packets sent to closed ports themselves (without forwarding them to the destination host), and are likely not to send back ICMP messages, due to security reasons. In Appendix~\ref{app:probing} we report that an estimated 85\% of the widely-used Internet hosts/firewalls do not send ICMP ``port unreachable'' messages.

\subsection{TCP Global States and Side Channels}
Our attack on NetBSD is TCP-based, and our attacks on Windows and OpenBSD IPv4 ID can also be used in TCP protocols. Therefore, we also survey TCP-based side channels, but only if they are relevant to any of these operating systems.

In 2005, Kohno et al. \cite{tcp-clockskew} demonstrated device identification based on the  device  clock  skew,  observed  in  the  TCP  timestamp  field. This work was later improved in \cite{hot-or-not}. Windows, OpenBSD and NetBSD do not implement TCP timestamps useful for the \cite{tcp-clockskew} and \cite{hot-or-not} attacks (Windows Server does not send a TCP timestamp at all, and OpenBSD and NetBSD have TCP timestamp at 0.5s resolution). In the past, TCP Timestamps could be used as a ``passive'' side channel for host alias resolution, as described in \cite{Bursztein2007TimeHS} \cite{misusing-TCP-timestamps}, but nowadays this only applies to macOS.

In 2010, Ensafi et al. \cite{10.5555/1929820.1929843} used model checking to discover information leakage due to the rate limit on a global TCP RST counter. Exploiting this requires the sender client to send the server either SYN packets for a closed port, or otherwise packets that purportedly belong to a non-existing TCP circuit. In both cases, a stateful firewall may silently drop the packets and thwart the attack. In Appendix~\ref{app:probing}, we report that an estimated 92\% of the widely-used Internet hosts/firewalls drop such packets.

Additional side channels are described by \cite{10.5555/1929820.1929843}, \cite{8279646} and \cite{Cao:2019:PUT:3319535.3354250}, which are based e.g. on TCP SYN cookies and TCP challenge ACKs. But these mechanisms are not employed (at least not by default) by Windows, OpenBSD and NetBSD.
We see, therefore, that for firewalled Windows Server, OpenBSD and NetBSD hosts, there are no currently known remotely exploitable TCP global states. 

\subsection{Host Alias Detection and De-NATting}
Early host alias detection and de-NATting relied on  global states that are trivial to observe and follow, and compared such global states between two endpoints. Bellovin's seminal 2002 de-NATting paper \cite{Bellovin2002} is based on a globally incrementing IPv4 ID as a global state. This concept was later applied to host alias resolution, e.g. with IPv6 ID \cite{10.1007/978-3-642-36516-4_16}. Client de-NATting based on the IPv4 ID of outbound DNS queries over UDP was described in \cite{Herzberg2017}. This attack relies on the incremental nature of the IP ID for a fixed $(\IP_{\src},\IP_{\dst})$ tuple, and thus cannot be applied to {\em server} de-NATting where a single host may use multiple source IP addresses.

Nowadays, IP IDs are not generated via a global counter (except for FreeBSD and NetBSD), and TCP timestamps are usually generated per TCP connection (except for macOS), thus these techniques are no longer in effect. Conceptually, it is still possible to mount host alias detection and de-NATting attacks using the more sophisticated global states described in the previous sub-sections, but here too, the severe limitations described above for these techniques apply. 
A host alias resolution technique based on packet delay sequences is described in \cite{cmc.2020.09850}, however this technique aims at {\em router} IP aliasing, which does not address the question of whether two IP addresses are served from a single host or from two distinct hosts on the same network.

\subsection{Container Co-residency Detection and Cross Container Information Leakage}

A container co-residency detection technique is described in \cite{10.1145/3411495.3421357}, but this attack is local, i.e. the attacker's container must co-reside with the target container. Our attack can {\em remotely} detect that two containers reside on the same host.

Cross-container leaks are reviewed in \cite{8023126}, but the attacks described there cannot be remotely mounted, whereas our attack is remote, and does not require the attacker to control a co-resident container.

\subsection{Other Side Channels and Exfiltration Techniques}
\begin{table}
\caption{Exfiltration from Networks}
\label{tab:exfil-networks}
\begin{center}
\resizebox{.5\textwidth}{!}{
\begin{tabular}{|l|l|l|l|}
\hline
\begin{tabular}[c]{@{}l@{}}Target\\ Network\end{tabular}       & \begin{tabular}[c]{@{}l@{}}Typical\\ Throughput\end{tabular} & \begin{tabular}[c]{@{}l@{}}Typical\\ Requirement\end{tabular}  & Examples                                              \\ \hline
Air-Gapped                                                     & 8b/h-4Kb/s                                                   & \begin{tabular}[c]{@{}l@{}}Proximity\\ (0.4m-30m)\end{tabular} & \cite{10.1007/978-3-319-66399-9_6},\cite{8514196}                                                    \\ \hline
Isolated                                                       & 1b/h-10Kb/h                                               & \begin{tabular}[c]{@{}l@{}}IP spoofing\\ (in some cases)\end{tabular} & \begin{tabular}[c]{@{}l@{}}\textbf{This paper}, \cite{hot-or-not}\end{tabular} \\ \hline
\begin{tabular}[c]{@{}l@{}}Regular\\ (Firewalled)\end{tabular} & \begin{tabular}[c]{@{}l@{}}100Kb/s and \\ above\end{tabular} & None                                                           & \cite{10.1007/978-0-387-72367-9_29},\cite{conf/pet/BrubakerHS14},\cite{220b1307a3014dbe96d0d59dfe07cecb},\cite{Kaminsky2004}                                             \\ \hline
\end{tabular}
}
\end{center}
\end{table}

Our exfiltration attacks are focused on isolated networks. Isolated networks are less secure than air-gapped (or nearly air-gapped) networks, but are more secure than ordinary networks. As can be expected, exfiltration techniques that work on less secure networks generally tend to have better throughput, but cannot be applied to more secure networks. Specifically, techniques that apply to less secure networks typically cannot be applied to isolated networks, and therefore cannot be compared, throughput-wise, to our techniques, that do work against isolated networks. Likewise, our exfiltration techniques do not work against air-gapped networks, or nearly air-gapped networks (networks behind a network pump). 

Exfiltration from air-gapped networks requires close proximity (0.4m-30m) \cite{190936}, which is a serious limitation. The throughput varies drastically among the techniques, from 8b/h to 4Kb/s \cite{8514196}. A typical  example is \cite{10.1007/978-3-319-66399-9_6} which describes an acoustical covert channel based on hard drive noise, yielding 3b/s throughput.

Our work is a rare example of exfiltration from an isolated network (without the proximity constraint). Another example is \cite{hot-or-not}, which is effective against Linux TCP services only, with 2b/h-8b/h throughput.

Finally there are examples of exfiltration techniques that do not apply to isolated networks. For example, \cite{10.1007/978-0-387-72367-9_29} uses a web counter as a covert channel, but this requires the DMZ target host to have (1) a shared web application between the internal and the external networks; and (2) to have a web counter in that application.  \cite{conf/pet/BrubakerHS14} uses public cloud service as a covert channel, and \cite{220b1307a3014dbe96d0d59dfe07cecb} uses multiple public services as a covert channel. All these covert channels are very unlikely to apply for isolated networks as their prerequisites are obviously inconsistent with network isolation.
In 2004, \cite{Kaminsky2004} demonstrated a covert channel via DNS queries (``DNS tunneling''). A throughput analysis can be found in \cite{DNS-tunneling-throughput}. Though it focuses on the downlink, it can be inferred that the uplink throughput is in the hundreds of Kb/s, many orders of magnitude better than our techniques. However, since DNS tunneling is well-known and well-understood for almost two decades, it is very unlikely to find isolated networks with unrestricted access to an Internet-connected DNS resolver. 

Table~\ref{tab:exfil-networks} compares exfiltration techniques for the three network types.

\section{Current Protocol Header Field Generation Algorithms}
\label{sec:current-protocols}
\subsection{Linux IPv4 ID for Connection-Less Transport Protocols}
\label{sec:linux-ipid-algo}
Starting with kernel v3.16, the Linux IPv4 ID generation algorithm for connection-less transport protocol packets (e.g. UDP, ICMP) uses an array $\beta$ of 2048 counters, and a corresponding array $\tau$ of their last access times. Algorithm~\ref{alg:linux-ipid} describes how an IPv4 ID is generated. Note that the time is measured in ``jiffies'' (a clock whose frequency is set at kernel compile time: $f$=250Hz by default, $f$=1000Hz for some MIPS platforms), $h$ is a hash function,  $proto$ is the IANA protocol number (1 for ICMP and 17 for UDP) and $net\_key$ adds a container-dependent quantity, if applicable. Also define $\textproc{random}(\emptyset)=0$. 

\begin{algorithm}
\caption{Linux IPv4 ID Generation for Connection-less Transport Protocols}
\label{alg:linux-ipid}
\begin{algorithmic}[1]
\Procedure{Generate-IPID}{}
\State $i \gets h(\IP_{DST},\IP_{SRC},proto,net\_key) \mod 2048$ 
\State $hop \gets 1+\Call{random}{\{0,\ldots,t_{now}-\tau[i]-1\}}$
\State $\beta[i] \gets (\beta[i]+hop) \mod 2^{16}$
\State $\tau[i] \gets t_{now}$
\State return $\beta[i]$
\EndProcedure
\end{algorithmic}
\end{algorithm}

\subsection{Windows Server IPv4 ID}
Note: since Windows is closed-source, extracting via reverse engineering, and documenting the Windows Server IPv4 ID generation logic is part of our contribution.

As of version 1903, 
Windows Server generates the IPv4 ID of outbound IPv4 packets as follows. Windows implements the IPv4 ID as a counter per each $(IP_{\src},\IP_{\dst})$ tuple that corresponds to outbound traffic, where $\IP_{\src}$ is an address on the local machine, and $\IP_{\dst}$ is a peer address. Such tuple-specific data is kept in an object called {\tt Path}. The set of {\tt Path} objects is maintained in a hash table called {\tt PathSet}. Windows keeps a {\tt PathSet} per compartment, i.e. one IPv4 {\tt PathSet} instance per container. A new {\tt Path} object is created upon the first attempt to send a packet from $\IP_{\src}$ to $\IP_{\dst}$, where there is no existing {\tt Path} object with these indices. When a new {\tt Path} object is created, the IPv4 ID field is initialized with random data. Then, for each packet sent from $\IP_{\src}$ to $\IP_{\dst}$, the IPv4 ID in the corresponding {\tt Path} object is incremented and used as ID.
Na{\"i}vely speaking, this setup provides no leakage, since each tuple runs its own independent IPv4 ID counter. However, an important aspect of the {\tt PathSet} life-cycle was left out so far -- how Windows handles the potentially almost infinite growth of {\tt PathSet}. Obviously {\tt PathSet} cannot be allowed to grow indefinitely, so {\tt PathSet} objects must occasionally be purged. Indeed, Windows removes objects from the {\tt PathSet} in batches, which we term {\em purge sequences}. Generally speaking, a purge sequence can be triggered either by {\tt PathSet}'s size exceeding some (pretty high) thresholds, or by a {\tt PathSet} growth of 5,000 or more new {\tt Path} objects in 0.5 second (``flood detection''). In both cases, a purge sequence is initiated, which goes over all the {\tt Path} objects in the {\tt PathSet} at a rate of 2000/sec and removes ``stale'' {\tt Path} objects -- objects that were last accessed over 60 seconds\footnote{10 seconds in Windows 10, determined by {\tt TcpipIsServerSKU}.} ago.

\subsection{macOS Server (Connection-Less Transport Protocols) and OpenBSD IPv4 ID}
Both macOS (for connection-less transport protocols only -- e.g. UDP and ICMPv4) and OpenBSD implement their IPv4 ID generation along the concept of keeping a list of the $\mathcal{M}$ recently used IPv4 ID values, picking a random ID value out of the remaining $65536-\mathcal{M}$ values, and updating the list of used values accordingly (popping the oldest value and pushing the freshly generated value).
For macOS, $\mathcal{M}=4096$, and for OpenBSD, $\mathcal{M}=32768$.
Thus, it is guaranteed that the IPv4 ID value is unique in intervals of length $\mathcal{M}$, at the cost of reducing the entropy from 16 bits to $\log_2 (65536-\mathcal{M})$ bits.

Note: the IPv4 ID in macOS is generated at the macOS kernel -- XNU. Therefore, the attack techniques we describe are actually inherent to XNU. However, macOS Server is the only widely-used {\em server} operating system based on the XNU kernel, therefore we name macOS as the attack target, rather than XNU.

\subsection{NetBSD TCP ISN}
A TCP Initial Sequence Number (ISN) is generated by each TCP connection party at the beginning of the connection, as a 32-bit random starting point for the sequence numbers it generates for the connection. A TCP client generates and sends ISN with the TCP SYN packet, and a TCP server generates and sends ISN with the TCP SYN+ACK packet. 

NetBSD generates the most significant 8 bits of the ISN as a sum of a 2Hz timer and a TCP connection counter, modulo 256. 

\section{Attacks}
\label{sec:attacks}

\begin{figure*}
  \subfloat[]{
	\begin{minipage}[t][1\width]{
	   0.24\textwidth}
	   \centering
	   \includegraphics[trim=8cm 2cm 14cm 0cm, clip=true, totalheight=0.25\textheight]{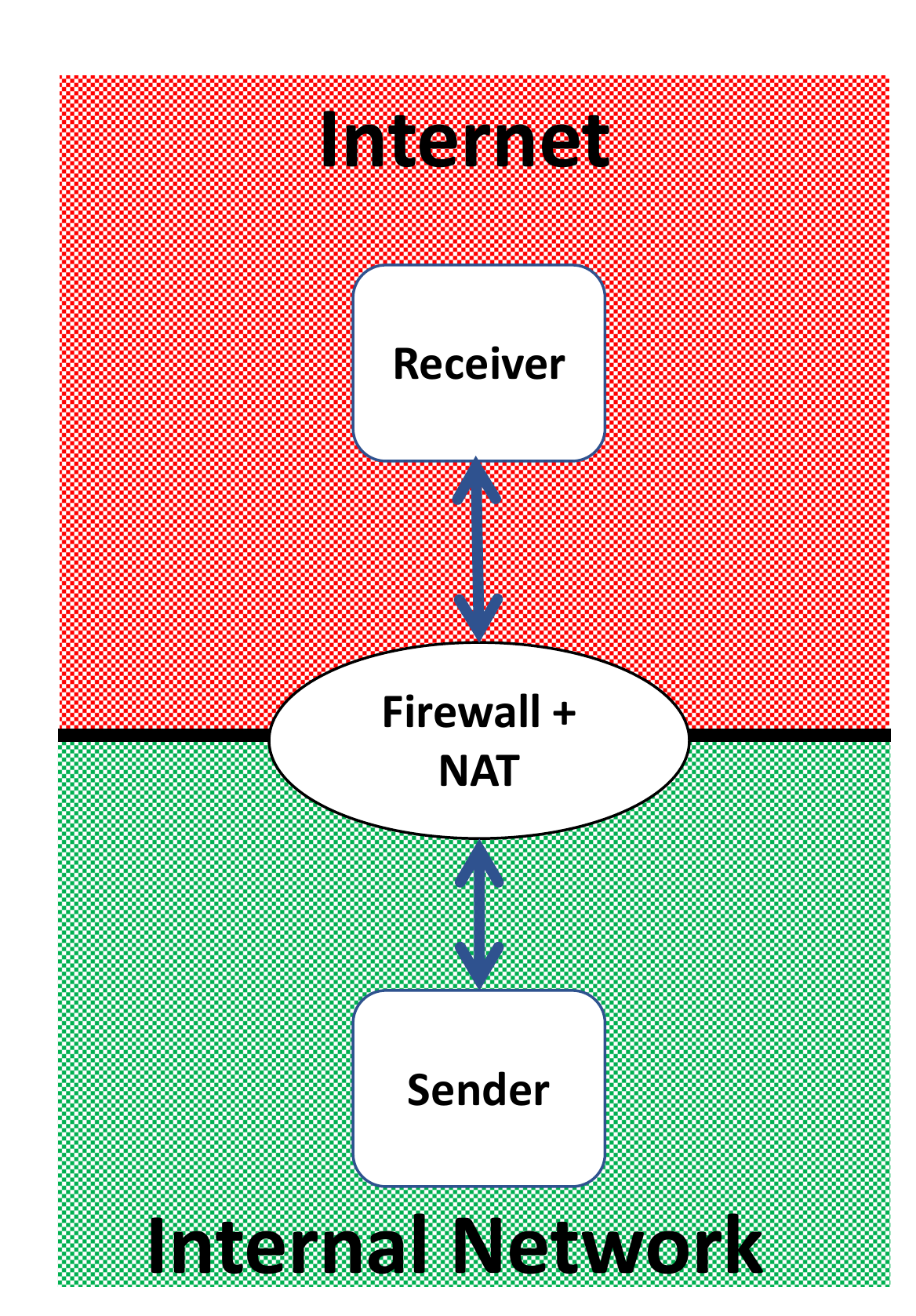}
	   \label{fig:piercing}
	\end{minipage}}
 \hfill 	
  \subfloat[]{
	\begin{minipage}[t][1\width]{
	   0.24\textwidth}
	   \centering
	   \includegraphics[trim=8cm 2cm 14cm 0cm, clip=true, totalheight=0.25\textheight]{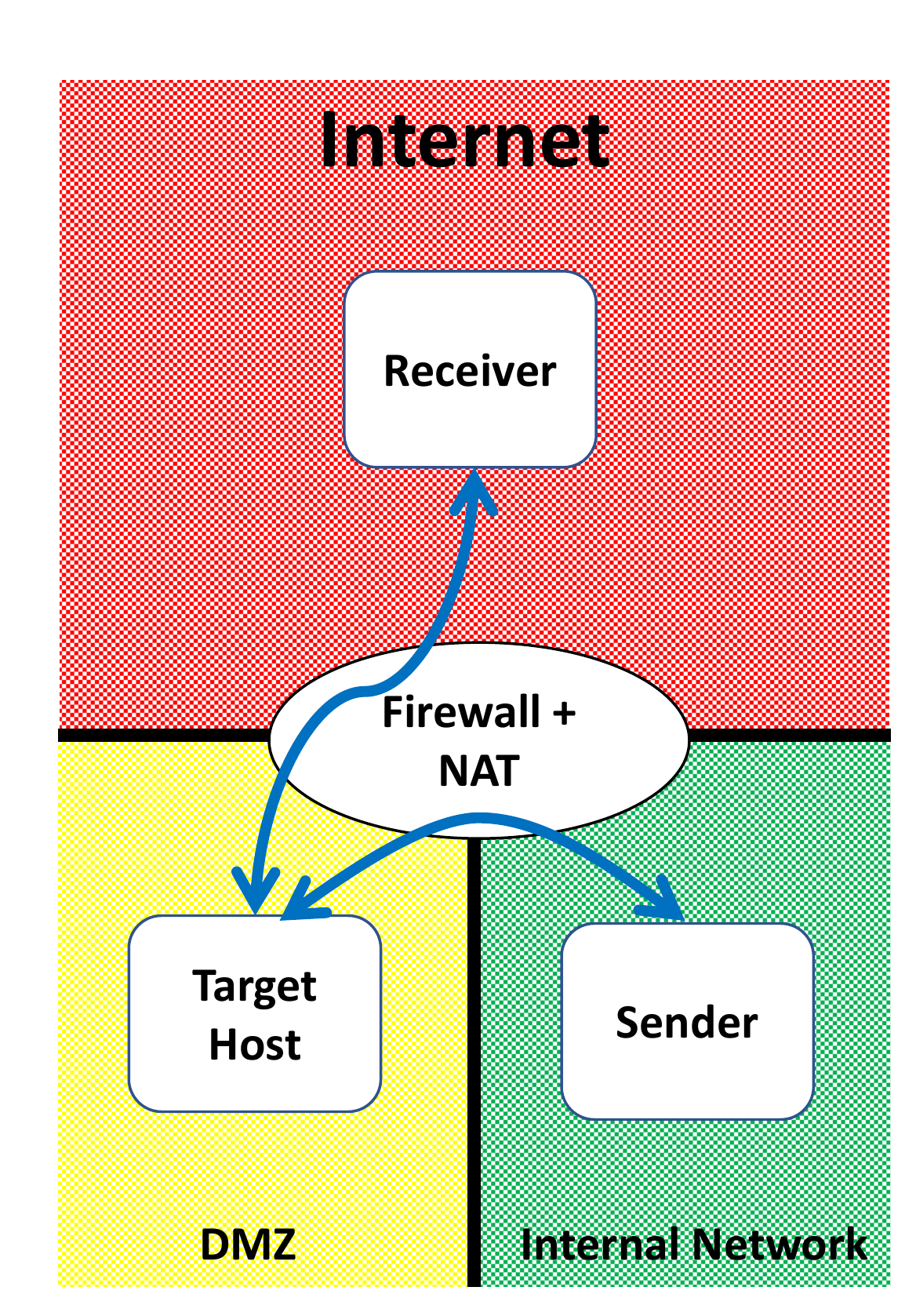}
	   \label{fig:exfiltration}
	\end{minipage}}
 \hfill	
  \subfloat[]{
	\begin{minipage}[t][1\width]{
	   0.24\textwidth}
	   \centering
	   \includegraphics[trim=8cm 2cm 14cm 0cm, clip=true, totalheight=0.25\textheight]{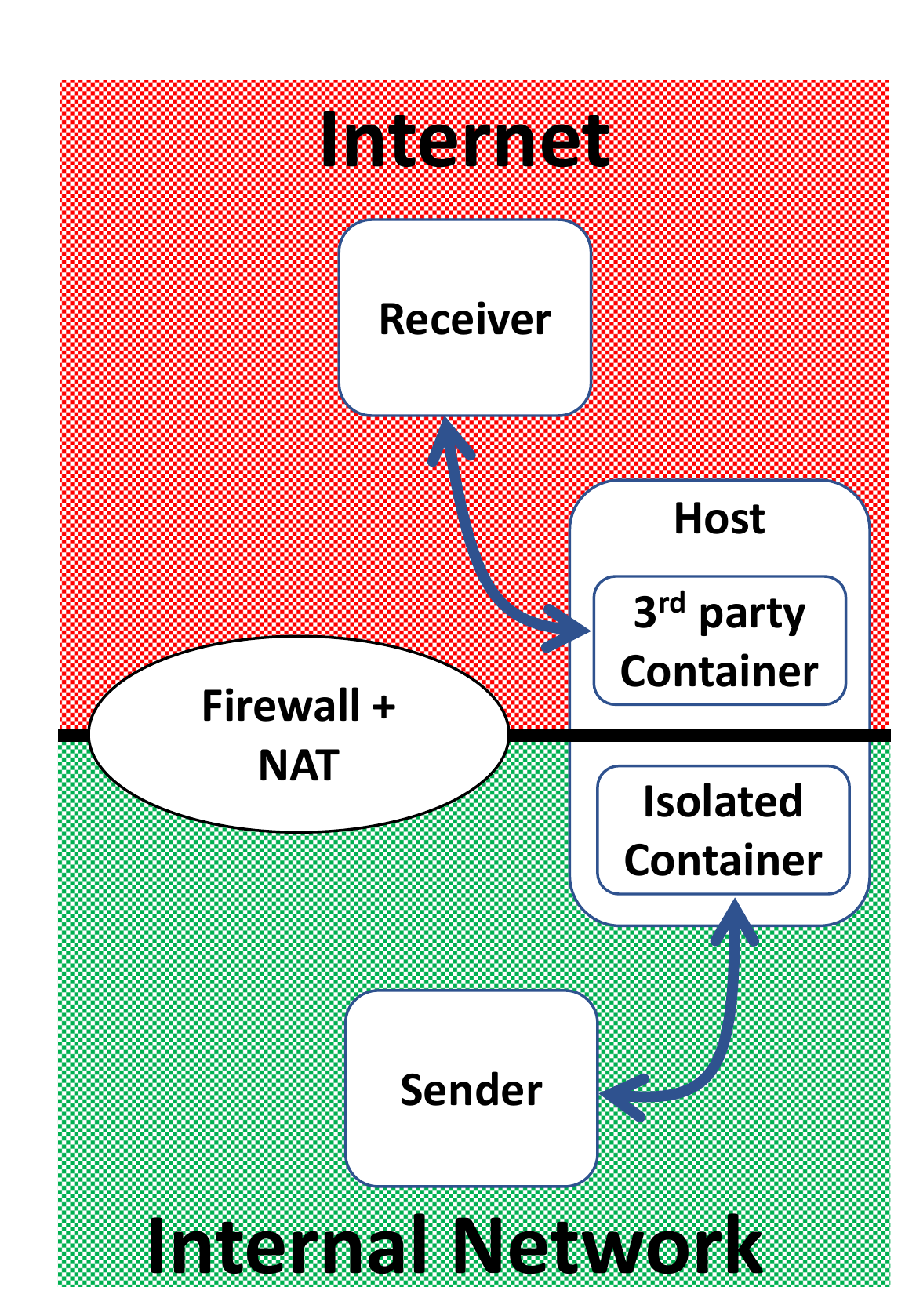}
	   \label{fig:exfiltration-containers}
	\end{minipage}}
\hfill	
  \subfloat[]{
	\begin{minipage}[t][1\width]{
	   0.24\textwidth}
	   \centering
	   \includegraphics[trim=8cm 2cm 14cm 0cm, clip=true, totalheight=0.25\textheight]{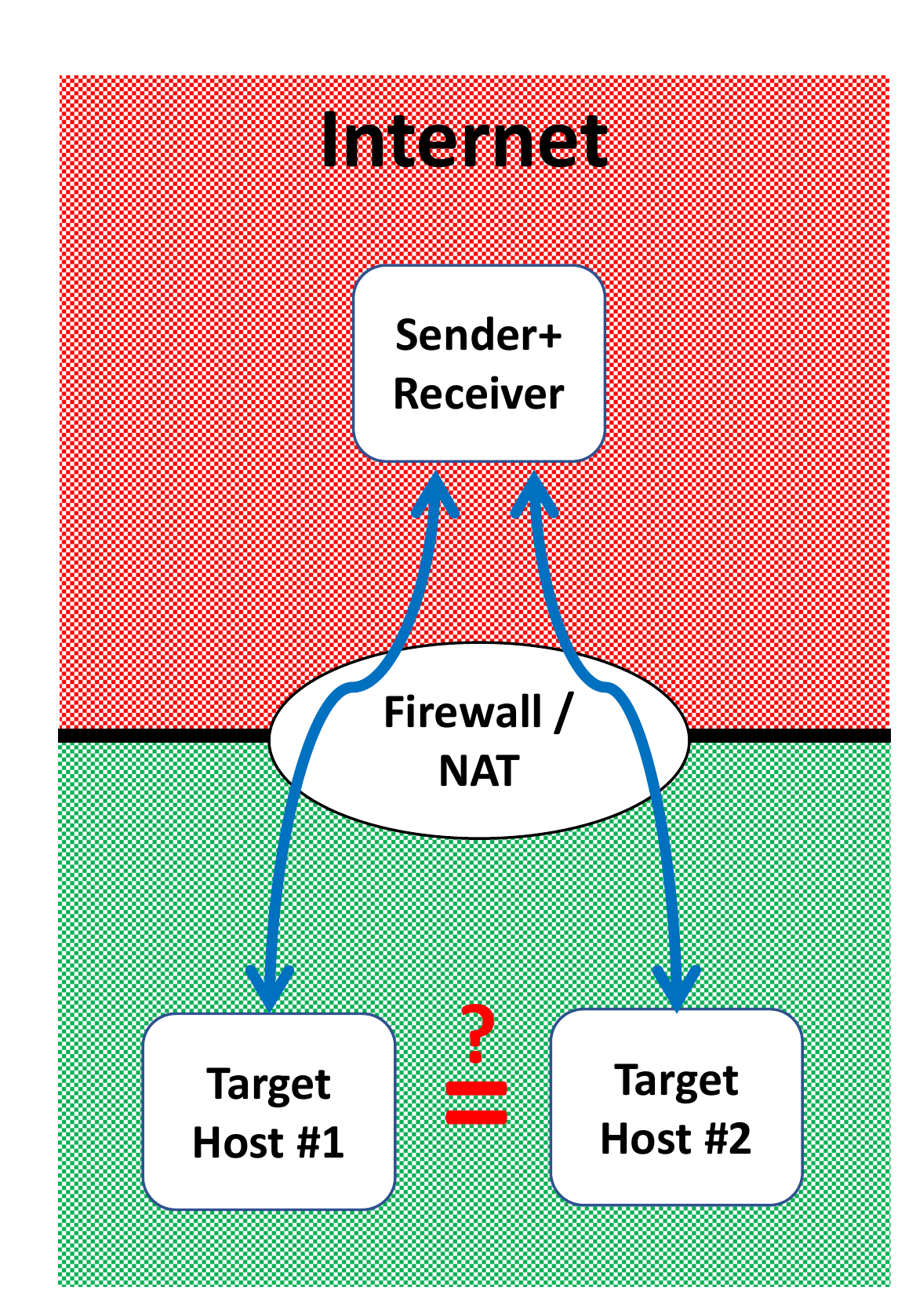}
	   \label{fig:host-alias}
	\end{minipage}}
	\captionsetup{justification=centering}
\caption{Network Diagram for (a) Firewall Piercing Exfiltration; (b) Exfiltration (via DMZ); (c) Exfiltration between Co-Resident Containers; (d) Host Alias Detection and de-NATting}
\end{figure*}

\subsection{Threat Model}
All our attacks are built around the same concept. In our attacks, there is a sender, a receiver and a target host (or hosts), identified by one or more endpoint designations, where an endpoint is defined as a combination of  IP address, protocol -- TCP, UDP or ICMP, and port (for TCP and UDP only). The attacker controls the sender and the receiver entities, but has no control whatsoever over the target host(s). Therefore, one of the main concerns of this research is to determine what attacks can be mounted in various implementations (operating systems) and roles of the target host. Thus, when we refer to an attack against an operating system, it is a shorthand to an attack against a target host running that operating system.

In one use case, the target host is a firewall preventing the sender and receiver from communicating directly. In other use cases, the target host is firewalled. The receiver and sender cooperate to fulfill their task, and are furthermore synchronized to a sub-second accuracy. 

We use a covert channel through the target host wherein its global state can be read and altered by sending and receiving packets, as described below. In Section~\ref{sec:channel-details} we will see that such a covert channel exists for many target host operating systems, though the covert channel details differ between the operating systems.

The sender entity signals a single bit by sending packets to one of the target host's designated endpoints. In most of our attacks, the sender can be an non-privileged process at the system it runs on, sending conventional packets. In fact, for web protocols, such traffic can be emitted by an HTML page rendered by a browser. However, in the Windows and Linux attacks, the sender must be able to spoof the source IP address of its packets, thus the sender must run as a privileged process on its machine. 

The receiver entity reads this bit by sending packets to one of the target host's designated endpoints (not necessarily the one used by the sender), receiving packets from the target and analyzing them. The receiver entity typically needs access to the raw packets at the IP level, and as such may need to run as a privileged process on its system. This is a reasonable assumption since the receiver is an attacker-controlled machine on the Internet.

In general, the attack proceeds by having the receiver send and receive some packets, thus recording the global state of the target host. Then the sender either sends packets to the target host (thereby changing its global state) to signal the bit ``1'', or does nothing (thereby retaining the global state in the target host) to signal the bit ``0''. Finally, the receiver reads the global state again, and determines whether it changed or not, thereby reading ``1'' or ``0'' respectively.

Naturally, there are timing constraints involved. For example, the receiver must conclude its first global state extraction before the sender is allowed to send its packets, and the sender must conclude its global state modification (if it takes place) before the receiver is allowed to send packets for its second global state extraction. Additional timing constraints are required for Windows covert channels.

Our attacks work best against network-wise idle (or almost idle) target hosts. Organic traffic may interfere with our techniques, although a minimal amount of traffic can be easily mitigated by adding some margins to the threshold parameters. For example, our Linux technique (Section~\ref{sec:linux-exfil}) expects an increase of no more than $\Delta t \cdot f$ in some counter when the machine is idle. When some traffic happens to use that counter, it is incremented further, say by up to $n$ packets per $\Delta t$ time slice. In such case, we choose $M>n+\Delta t \cdot f$ (instead of $M>\Delta t \cdot f$). A mild amount of traffic can be mitigated by introducing redundancy (e.g. majority over three same-bit transmissions) and/or error correction codes (e.g. Reed-Solomon \cite{Reed1960PolynomialCO}).

We demonstrate our UDP attacks using the HTTP/3 protocol. Specifically, the sender sends QUIC Initial packets \cite{rfc9000} to the target host, and the receiver sends Initial packets and receives QUIC response packets. We demonstrate our TCP attacks using the HTTP/1.1 and SSH protocols. In both cases, the attack uses open UDP/TCP port on the target device. In our ICMP attacks, we use the standard, built-in ICMP Echo Request-Reply mechanism, which is commonly open\footnote{We do not consider opening ICMP to the Internet a security vulnerability. The security issue is with the  implementation of the underlying IP layer.} to the Internet. In Appendix~\ref{app:probing} we report that 89\% of the widely-used Internet hosts/firewalls surveyed responded to ICMP Echo requests.

In the firewall bypassing scenarios, we focused on data exfiltration attacks, but in general, the opposite direction -- sending data into an internal network -- is also possible. This can facilitate e.g. a Command \& Control channel for the malware. We demonstrated this direction in one test (LF-1).

\subsection{Source Address Spoofing and Sender Privileges}
For the Linux and Windows attacks, the sender needs to be able to spoof the source IP address of its packets. This requires the sender to have root privileges on a machine which is on an internal network. With root privileges, spoofing is typically trivial, using e.g. raw sockets or libpcap. The firewall's SAV protection still allows a device on the internal network (the sender) to send packets with a source address from the same (internal) network to the firewall or to the DMZ. This suffices for our firewall subversion and exfiltration attacks. For the host alias resolution attacks where the sender is on the Internet, SAV employed by ISPs may prevent address spoofing. According to \cite[Fig. 4]{Spoofer}, 11.9\%-30.5\% of the ASes do not employ such filtering on their customers ({\em outbound} filtering),  hence  can  be  used  by  the  attacker  to  send  spoofed packets from. Naturally, the attacker is free to choose such a network, thus having enough non-filtering networks guarantees that the attacker can mount the attack.
We conclude therefore that packet spoofing over the Internet is feasible, hence the Linux and Windows host alias resolution attacks over the Internet are practical.

\subsection{Firewall Piercing Exfiltration}

This use case places minimal requirements on the network topology. The attacker's goal is to exfiltrate information from a compromised machine (the sender) in a well guarded enterprise network. It is assumed that the sender cannot send packets to the Internet. Fig.~\ref{fig:piercing} illustrates the network topology (each network has its own background color, blue arrows represent direct communication, bold black lines define network boundaries enforced by the firewall).

The attacker uses an ICMP-based covert channel to exfiltrate one bit at a time from the sender to the receiver. The only requirement is that the firewall responds to ICMP Echo requests both from the Internet and from the internal network.

\subsection{Firewall Subversion and Exfiltration (via DMZ)}
\label{sec:firewall-subversion-exfil}

In this use case, the attacker's goal is to exfiltrate information from a compromised machine (the sender) in a well guarded enterprise network. It is assumed that the sender cannot send packets to the Internet directly. Fig.~\ref{fig:exfiltration} illustrates the network topology. The sender can send and receive packets to/from a server host in a DMZ segment of the enterprise network (the target host). This server serves both internal clients such as the sender, and external Internet clients such as the receiver. The target host is protected by a stateful firewall both from the internal network and from the Internet. The firewall also enforces SAV between the three networks (internal network, DMZ and Internet).

The attack uses the covert channel to exfiltrate one bit from the compromised machine (the sender) through the target host (the DMZ server) to the attacker machine on the Internet (the receiver). 

\subsection{Exfitration via Co-Resident Containers}

A special case of the exfiltration technique described in Section~\ref{sec:firewall-subversion-exfil} is exfiltration between containers. Fig.~\ref{fig:exfiltration-containers} illustrates this attack scenario. In this case, there are two co-resident containers on the same host (kernel), which are supposedly completely isolated. The ``isolated container'' resides in a completely isolated local network, shown as the green network in Fig.~\ref{fig:exfiltration-containers}. The other container, depicted as ``\nth{3} party container'' can belong to any entity, and is only required to run an Internet-facing service in the protocol designated for exfiltration. None of these containers is assumed to be controlled by the attacker. 

A malicious sender that resides on the local network can exfiltrate data to an Internet receiver using the same technique described in Section~\ref{sec:firewall-subversion-exfil}, assuming the global state is shared between the containers.

In a variation of this use case (not depicted in Fig.~\ref{fig:exfiltration-containers}), the sender may be malware running on the isolated container, without root privileges. Such a sender can use 127.x.y.z as local address destinations, instead of spoofing packet source addresses.

\subsection{Host Alias Resolution, De-NATting and Co-Residency Detection}

In this use case, the attacker's goal is to determine whether two endpoints reside on the same host (even across containers, if possible) or not. The attacker is an Internet machine, and the endpoints are located behind a stateful firewall. The attacker runs both the sender logic and the receiver logic, thus there is no importance in distinguishing between the receiver and the sender, and we will use the term ``attacker'' for both. The attack applies the receiver logic to one endpoint, and the sender logic, specifically writing the bit ``1'', to the other endpoint. If both endpoints reside on the same host, the receiver will read the bit ``1'', but if the two endpoints reside on different hosts, then writing to one host does not affect the other host, and the receiver will read the bit ``0''. This use case is illustrated in Fig.~\ref{fig:host-alias}.
If the specific covert channel works across containers, then so will this attack.

Since the attack implementations for host alias resolution and de-NATting are identical, we will henceforth use the term ``host alias resolution'' to denote both, and we will note whether the attack also applies to container co-residency detection, per case.

\section{Information Leakage and Covert Channels: Global States in Protocol Fields}
\label{sec:channel-details}
We now show the existence of information leakage in the ICMP/IPv4 and UDP/IPv4 implementations of some popular server operating systems, and in the TCP implementation of NetBSD. The information leakage enables the receiver -- a client of the server -- to obtain information on the state of the server. We then show how to exploit this weakness to construct a covert channel, in which a sender client, typically running on a different machine than the receiver, signals a bit by forcing the server state to meet a certain condition or fail to meet the said condition. Thus, the receiver and sender engage in a covert channel protocol in which the sender sends an arbitrary bit to the receiver. Our attacks share some common principles, but differ in their technical details between operating systems. In some cases, our techniques can also be applied to TCP/IPv4. 

\subsection{Linux Connection-Less IPv4 ID}
\label{sec:linux-exfil}
The global state that enables a covert channel is the IP ID counter used by the target host to send packets to the receiver, i.e. $\beta[R]$ (defined in Section~\ref{sec:linux-ipid-algo}) where
$$R=$$
$$h(\IP_\receiver, \IP_\target, proto, net\_key) \mod 2048$$

The sender needs to send packets to the target rapidly. 
When the sender can force the target to send more than $f$ packets per second from $\beta[R]$, the receiver will be able to detect this by sampling packets generated by the target host using the $\beta[R]$ counter, and observe that $\beta[R]$ incremented at a rate $>f$. Whereas when the target host does not receive sender packets, $\beta[R]$ will increment at a rate $\leq f$.

There is a problem with the above approach, though. The probability of the sender's IP address hash to collide with $R$ is $\frac{1}{2048}$. The solution is simple: note that the sender needs only to force the target host to send packets whose IP ID is generated using $\beta[R]$. The sender does not need to actually receive and read those packets. And so, the sender can simply craft packets that elicit the target host's response, spoof the source address of these packets (to an unused internal network IP address), go over enough source addresses to practically guarantee that one of the responses uses $\beta[R]$ (as explained below), and thereby fulfill the condition for the attack.

We now expand the above technique into a covert channel. For simplicity, we assume that the receiver owns a single Internet IP address, however practically, we expect the attack to be mounted with a receiver that owns multiple IP addresses. Assume the receiver can sample the target host at $\Delta t$ intervals, i.e. at a frequency of $\frac{1}{\Delta t}$. Denote by $M$ an attack parameter (threshold) $\Delta t \cdot f \leq M \ll 2^{16}$. Ideally, $M$ should be minimal to reduce the required bandwidth of both parties (and also to avoid wrapping around), yet should take into consideration possible noise and jitter, i.e. $M$ should be slightly higher than $\Delta t \cdot f$. The sender chooses a list of IP addresses that will be used for spoofing. The list length $L$ is a function of the acceptable probability $p$ of bit read error (for bit ``1''), which happens when all $L$ addresses do not elicit an answer that uses $\beta[R]$. We choose a parameter $\Lambda$, such that $p=(1-\frac{1}{2048})^\Lambda \approx e^{-\frac{\Lambda}{2048}} < 0.001$, i.e. $\Lambda \geq 14148$. For a receiver with a single IP address, we set $L=\Lambda$ (in our experiments, we rounded this down to $L=14000$). Assume the sender can send $B$ packets per second. Then transmitting one bit takes $\frac{(2M-1) \cdot L}{B}$ seconds. The attack works as follows:
\begin{itemize}
    \item Receiver: During the transmission time of a single bit, i.e. for $\frac{(2M-1) \cdot L}{B}$ seconds, the receiver samples the target host at intervals $\Delta t$ and records the IPv4 ID field of the server packets -- $\ID_i$. It then checks, for each $i>0$, if $(\ID_i-\ID_{i-1}) \mod 2^{16} \geq M+1$. If this is met for any $i>0$, then the bit transmitted is determined to be ``1''. Otherwise the bit is determined to be ``0''. We show below that it is guaranteed that the sender increments the IP ID counter by at least $M$ for one interval, and the receiver's first sample increments it at least by 1, hence a signalled interval should have an increment of at least $M+1$.
    \item Sender: To send a bit ``1'', the sender goes through the $L$ addresses, and for each, spoofs $2M-1$ packets originating from the address and whose destination is the target host, during a $\Delta t$ time interval, over ICMP or the UDP protocol of choice. To send a bit ``0'', the sender does nothing.
\end{itemize}
The sender needs to send $\frac{2M-1}{\Delta t}$ packets per second from each spoofed address, since there is no expectation to synchronize the time the sender's packets arrive at the target host, with the time the receiver's packets arrive at the host. All $2M-1$ packets from the same spoofed address should be sent in rapid succession. By sending $2M-1$ packets at an arbitrary $\Delta t$ interval, the sender ensures that at least one $\Delta t$ interval sampled by the receiver will have at least $M$ sender packets in it. In other words, the sender can send a burst of $2M-1$ packets from one spoofed address (assuming this takes less than $\Delta t$) and move on immediately to the next spoofed address.

Note that for this technique to yield correct results, no other network traffic emitted by the target host, other than to the sender and to the receiver, must be using the $\beta[R]$ counter. This is a reasonable assumption, since the probability of the host's outbound traffic to a random address to use $\beta[R]$ is $\frac{1}{2048}$.

This covert channel works across containers. Even though the $net\_key$ introduces dependency in the container for which the hash is calculated, the $\beta$ table itself is shared among all containers. Since the attack does not rely on having the same hash function for the receiver and the sender, it follows that the attack covers the case wherein the receiver is served by one container, and the sender is served by another container.

The covert channel also works across protocols ($protocol$=1 for ICMP, 17 for UDP), by the above argument.

Ideally, the attack is carried out with a receiver that owns $n \gg 1$ IPv4 addresses. In such a case the sender needs to spoof only $L=\frac{\Lambda}{n}$ IPv4 addresses (similar to the birthday paradox), and thus the attack speeds up by a factor of $n$. This requires the receiver to send an aggregate of $\frac{n}{\Delta t}$ packets per second to the host. The receiver needs to first ensure that each IP address it owns uses an exclusive counter. For this, the receiver can employ the techniques described in \cite[Section 6]{KP19-usenix} to detect collisions in pairs of IP addresses, and remove one address from each such pair, which results in a collision-free list. Also, once the receiver knows which IP addresses have counters that collide with those of the sender's addresses, the receiver can send packets only to these IP addresses, which saves bandwidth. Moreover, for the firewall piercing attack, the sender (assuming it is on the internal network) cannot spoof too many IP addresses, since the addresses it spoofs are on the firewall's local network\footnote{This restriction does not apply to other attack scenarios, since there, the sender and the target host reside on different networks.}, and thus the firewall has to have them in its ARP cache, which is limited in size. As long as there are less than {\tt gc\_thrash1} (defaults to 128) ARP entries, garbage collection will not run \cite{arp-linux}, therefore the sender must keep the number of spoofed addresses below 128 (e.g. $L=127$). This is feasible as explained above when the receiver can listen on $n\geq \lceil \frac{\Lambda}{127} \rceil$ IP addresses.

The limiting factors in this attack are the sender's bandwidth (vs. $M\cdot L$), and the network jitter between the receiver and the host (vs. $\Delta t$) as a jitter value close to $\Delta t$ may cause the receiver's probe packets to arrive at the host out of order.

See Appendix~\ref{app:improved-attack-Linux} for an improved attack variant wherein feedback from an opposite direction channel is used to significantly reduce the traffic, bandwidth, and number of addresses required for the channel.

Linux generates the IPv6 Identification header extension field in a very similar manner, and using the same table. IPv6 fragmentation attacks are out of scope for this research. 
\subsection{Windows Server IPv4 ID}
\subsubsection{Information Leakage}
Suppose at time $t$, a receiver at IPv4 address $\IP_1$ received a packet from a Windows Server target host at $\IP_2$, with IPv4 identifier $\ID$, and at time $t''>t$, the receiver received a second packet from the target host, with IPv4 identifier $\ID''$.
$\ID$ originated from a {\tt Path} object indexed by ($\IP_1$,$\IP_2$), at time $t$. $\ID''$ originated from a {\tt Path} object indexed by ($\IP_1$,$\IP_2$), at time $t''$. If the {\tt Path} object indexed by ($\IP_1$,$\IP_2$) remained intact in the time window $[t,t'']$ then $\ID''=\ID+1 \mod 2^{16}$. However, if, at some time $t'$, $t<t'<t''$, this  {\tt Path} object was purged, then the second packet originated from a freshly created {\tt Path} object, whose IPv4 ID is initialized with a random value, i.e. $\ID''=\ID+1 \mod 2^{16}$ with probability $2^{-16}$. The leakage then is whether a purge occurred at time $t'$ or not.

The purge timing is essential for the leakage to occur. One restriction is $t'-t\ge60s$, as only {\tt Path} objects whose last access time is over 60 seconds ago are purged. Another restriction relates to the exact purge timing of the {\tt Path} object. As we will see below, there may be up to 20000 {\tt Path} objects in the {\tt PathSet} (or even a bit higher, if the target host is not completely idle). Since the purge rate is 2000/sec, it may take up to 10 seconds for the specific {\tt Path} object to get purged. Therefore, the additional condition is $t''-t'\ge10s$.

From the above $t''-t\ge70s$, i.e. the minimal time between receiver samplings is 70 seconds. We also need take into account the fact that the flood detection triggering consumes another second, i.e. the probes have to be 71s apart. Practically, we need to pad this a bit to accommodate for {\tt Path} objects that organically get inserted into {\tt PathSet} during the normal system run. An additional 1s accommodates 2000 such objects.

Interestingly, due to a bug in Windows, oftentimes there is an additional purge sequence right after the first purge sequence, taking additional 5 seconds in our use case. This additional purge sequence does not interfere with our covert channel, because if it extends a few seconds beyond the next probe, that probe will not be affected, as it is not stale.

It should be noted that a leakage does not occur across containers (compartments), since each compartment maintains its own IPv4 {\tt PathSet}.

\subsubsection{Covert Channel}
The receiver and the sender have synchronized transmission cycles. 
The receiver samples the IPv4 ID of the host at the beginning and at the end of every transmission cycle, by sending $K$ UDP requests that elicit UDP answers from the host, and observing the IPv4 ID of the host packets. Multiple requests are needed to cater for packet loss. The receiver determines that the transmitted bit is ``0'' if the difference (modulo 65536) between the IDs is less than the number of packets sent to the host during the cycle -- $2K$. Otherwise, the bit is determined to be ``1''. The choice of $K$ can be arbitrary, as long as the value is very small, with very little impact on the channel's performance.

The sender signals a bit ``0'' by not doing anything during the transmission cycle. This guarantees that the {\tt Path} object in the host, that belongs to the receiver, is not purged, and therefore the IPv4 IDs will remain sequential for the end-of-cycle probe.
On the other hand, for the bit ``1'', the sender needs to force that {\tt Path} object to get purged. In the end-of-cycle probe, the receiver will be assigned a new {\tt Path} object, with a random IPv4 ID base, which is very likely to fall outside the range for bit ``0''.

Thus, to transmit the bit ``1'', the sender needs to inflict a purge cycle on the target host. For this, the sender can trigger the {\tt PathSet} flood detection, by forcing the target host to send packets to 10000 different IP addresses in a single second. Since the flood detection is measured in intervals of 0.5 seconds (5000 new {\tt Path} objects per 0.5 second), this ensures that flood detection will be triggered. In the QUIC use case, the sender can achieve this by sending QUIC Initial packets (1228 bytes at the IP layer) from spoofed IP addresses to a UDP port on the target host listening to a protocol that is on top of QUIC, e.g. HTTP/3. The target host is then obliged to send a QUIC response to the IP address from which the packet allegedly originated. The required bandwidth is therefore 98.24Mb/sec, well below the limit of modern WiFi and Ethernet networks, and an achievable upload speed by fiber Internet connections.

This attack's performance is conditional upon the sender having bandwidth of at least 10000 packets/s in the chosen protocol, and takes at least 71s per bit. 

A ``1'' bit can be interpreted as a ``0'' when the new {\tt Path} object contains an IP ID that accidentally falls into the range allowed for bit ``0''. This happens with probability $\nicefrac{2K}{65536}$, and only for ``1'' bits. Therefore, the bit error rate is 
$\nicefrac{K}{65536}$.

Note that this attack is not disruptive. Organic {\tt Path} objects are indeed removed from the {\tt PathSet}, but only if their last access time was over 60 seconds ago. This means that two IPv4 packets/fragments generated over 60 seconds apart may have the same IPv4 ID, with probability $\nicefrac{1}{65536}$. This is not an issue since it is highly unlikely that the first fragment will not be reassembled at the destination for this long. 

An Internet attack cannot be mounted over ICMP, because Windows only responds to ICMP Echo requests from the local network.

\subsection{macOS Server (Connection-Less Transport Protocols) and OpenBSD IPv4 ID}
For two IPv4 ID values $\ID_0, \ID_m$ sampled $m>0$ packets apart, we have (recall that $\mathcal{M}$ is the uniqueness interval):
\small
\[ 
\Pr\{\ID_m=\ID_0\}= \left\{
\begin{array}{ll}
      0 & m \leq \mathcal{M}\\
      \nicefrac{1}{(65536+m-2\mathcal{M}-1)} & \mathcal{M}<m \leq 2\mathcal{M} \\
      \approx \nicefrac{1}{65536} & 2\mathcal{M} < m \\
\end{array} 
\right. 
\]
\normalsize
Thus when we sample two IPv4 ID values, we can statistically distinguish between a case where $m' \geq \mathcal{M}$ packets were generated in between (in such case, 
$\Pr\{\ID_{m'+1}=\ID_0\} \geq \nicefrac{1}{65536}$) and a case where $m'< \mathcal{M}$ packets were generated ($\Pr\{\ID_{m'+1}=\ID_0\}=0$). Practically, the receiver first gets $K$ packets from the target host, as rapidly as possible. Then the sender forces the target host to send $\mathcal{M}$ packets to signal the bit 1, and no packets to signal the bit 0, and finally the receiver gets additional $K$ packets. $K$ should be picked such that $K^2 \gg 65536$ and $2K < \mathcal{M}$ (ideally $K \ll \mathcal{M}$). 
The receiver now checks whether there are any collisions between the two sets. When a bit ``0'' is transmitted, we clearly have $\Pr\{C=0\}=1$. When a bit ``1'' is transmitted, the probability of {\em no} collisions is:
$$\Pr\{C=0\}=\prod_{0 \leq i,j <K} (1-\Pr\{ID_{j+\mathcal{M}+K}=ID_i\})$$
$$\leq \prod_{0 \leq i,j <K} (1-\nicefrac{1}{65536}) \approx e^{-\frac{K^2}{65536}}$$
Since $\frac{K^2}{65536} \gg 1$ this probability is negligible. 

To summarize, we described a covert channel, in which the receiver forces the target host to send back $K$ packets, then the sender forces the target host to send back $\mathcal{M}$ packets (or none at all), and finally the receiver forces the target host to send back additional $K$ packets. The receiver looks for collisions in the IPv4 IDs of the two $K$ packet sets. If collisions are found, the receiver infers that the sender interacted with the target host (bit ``1''). If no collisions are found, the receiver infers the sender did not interact with the target host (bit ``0'').

The limiting factors for this attack is the receiver's bandwidth (vs. $K$) and the sender's bandwidth (vs. $\mathcal{M}$). The jitter (assuming low values) has almost no significance here, as the order of packet arrival is unimportant neither for the sender, nor for the receiver. And since both the sender and the receiver are required to send thousands of packets, which takes a lot more time than the typical jitter, the bandwidth overshadows the jitter as a dominant factor.

See Appendix~\ref{app:improved-attack-OpenBSD} for a stealthier variant of this attack.

In macOS 11.3 and above, an ICMP Echo rate limit is imposed, with a threshold value in the range 251-500 chosen at random once per boot, and applied to 1-2 second intervals. When the rate limit is exceeded in an interval, the ICMP Echo response is generated with probability inversely proportional to the excess ICMP Echo request packets. That is, if the random rate limit is $R$ and $N>R$ ICMP Echo requests were already received in the interval then the next ICMP Echo request will be answered with probability $\frac{1}{N-R}$. The above attack can be mounted almost as-is by throttling it to under 125 packets/s. But a more efficient attack can simply take advantage of the rate limit to signal bits. To signal ``1'', the sender sends a burst of e.g. 600 ICMP Echo requests (which guarantees an excess of at least 100 requests, thus ensuring that subsequent ICMP Echo requests in that time period has a probability $\leq\frac{1}{100}$ to be answered by macOS). The sender does nothing for ``0''. The receiver then sends an ICMP Echo request to the host, which is always answered if the sender sent nothing (``0''), but is not answered (with probability $>0.99$) if the sender sent 600 requests. This procedure needs to be fortified with some error correction logic (to take care of the occasional response from the server even in the case the rate limit was exceeded). Thus a single bit can theoretically be signalled in two seconds. Due to logistic reasons, we could not test this attack. 

\subsection{NetBSD TCP ISN}
The TCP ISN -based global state is the number of TCP connections established since boot (modulo 256). 

The covert channel works as follows: the receiver sends a TCP SYN packet to the NetBSD target host, and records the most significant 8 bits of the ISN in the target host's SYN+ACK response -- $m$. The sender then sends two or more TCP SYN packets to the host, for bit ``1'', or no packets for bit ``0''. Finally, after $\Delta t$ time, the receiver sends a TCP SYN packet to the host, and records $m'$ from the host's SYN+ACK response. Then if $(m'-m) \mod 256 > 1+\lceil 2\Delta t \rceil$, there was another connection served by the target host, i.e. the sender signalled the bit ``1'', otherwise the sender signalled ``0''.

The attack requires that the jitter be much smaller than $\Delta t$.

Note that unlike the well known sequential IPv4 ID attack for NetBSD, our attack also works with TCP/IPv6, and can combine TCP/IPv6 and TCP/IPv4 senders and receivers. 

\subsection{Additional Attacks}
We also found additional new attacks against 
some of the above operating systems,
that are disruptive and/or inferior in performance (packets required) to the attacks described above. For sake of completeness, we describe these attacks in Appendix~\ref{app:more-attacks}. Note that if/when the attacks described above are addressed by 
the respective operating system vendors, 
the attacks in Appendix~\ref{app:more-attacks} may take their place as the most effective attacks (unless they are fixed as well).

\subsection{Practical Considerations}
\subsubsection{RTT}
Our attacks require some synchronization between the time of arrival of the sender packets and the receiver packets, at the target host. Interestingly, such synchronization is not affected by the RTT itself, assuming each party knows its RTT to the target host beforehand. For simplicity, let us assume a {\em fixed}, {\em symmetric} RTT. In order for a packet to arrive at time $t$ to the target host, the sending party needs to send it at time $t-\frac{RTT}{2}$, assuming the party's machine time itself is synchronized. In reality, RTT is not fixed, and the RTT variance, or jitter, determines the synchronization granularity. Likewise, in reality, $\frac{RTT}{2}$ is only an approximation to the end-to-end time between two parties -- the actual value may differ somewhat due to asymmetry in routing paths. This unknown quantity can also be accumulated into the synchronization granularity. 

Moreover, in our attacks, each packet sent by the sender and by the receiver is independent of the previous packets, i.e. the attacks are entirely non-adaptive. This means that neither party needs to wait for any response (to a previous packet) to arrive from the target host, before sending the next packet. Therefore, there is no significance to the RTT in timing the attacks. The only exception is some TCP/IP attack scenarios wherein one or two parties send an RST in response to the host's SYN+ACK answer. The RST packet contains the host-generated TCP sequence number from the host's SYN+ACK, and thus it relies on a previous packet from the host, i.e. subject to RTT. However, sending the RST packets can be done in parallel with the attack's main packet stream, and thus the RTT does not affect the overall attack time in this case as well.

\subsubsection{Jitter and Bandwidth}
The limiting factor for the attack speed is the sender's bandwidth (which we throttle for some attacks) and the jitter, which is typically at milliseconds granularity. In fact, due to its typically low value, jitter is a limiting factor only in Linux and NetBSD, and the sender bandwidth is a limiting factor in Linux, OpenBSD and macOS. In Windows, the attack time is dominated by the 60s time-to-live limit imposed on stale {\tt Path} objects, and is thus unaffected by jitter and bandwidth (except that it requires the sender to be able to send 10000 packets/s bursts).

\subsubsection{Synchronization between the Sender and the Receiver}
We assume that the sender and receiver machines are synchronized at 0.1s resolution. This is a fair assumption, given the accuracy of present day hardware clocks, combined with NTP which keeps the clocks from drifting \cite{NTP-FAQ}. Thus the sender and receiver can decide on a UTC time (in seconds) when they begin the exfiltration session. Since the exfiltration techniques are based on the receiver polling the state before and after the sender, the receiver can poll 0.1s before the bit transmission start time, and poll again 0.1s after the bit transmission end time (all times are in the target's time frame, i.e. both sender and receiver should compensate based on their respective RTTs, by transmitting at their $t-\frac{RTT}{2}$). This compensates up to 0.1s synchronization offset between the sender and the receiver. A coarser-grained synchronization (e.g. 0.2s) can be accommodated via increasing the bit transmission time by 0.2s.

\subsubsection{Interference (Organic Outbound Traffic from the Target Host)}
In the firewall piercing scenario, using ICMP, the firewall can be considered idle in terms of network traffic, since the vast majority of the traffic it handles is as a router, where the firewall is not required to generate e.g. IPv4 IDs. 

In general, our techniques are not sensitive to a mild level of interference. The macOS and OpenBSD techniques only require that the organic outbound traffic bandwidth (in terms of packets/s) be lower than the bandwidth generated by the sender (or more precisely -- lower that the threshold that the reader interprets as a signalled bit ``1''). Raising this threshold can compensate for interference. The Windows technique can easily compensate several thousands IP tuples (our experiment setup actually already has such compensation in place) by allowing a bit more time for purging them (purging 2000 tuples takes 1s). The Linux technique is affected by interference only if the specific bucket used for the communication happens to also serve organic traffic, which is very unlikely (probability $\frac{1}{2048}$ per destination IP address).

\subsubsection{Packet Loss}
Our techniques are not sensitive to a mild level of packet loss. In general, the attacks need a certain number of packets, or a certain rate of packets in order to succeed. Packet loss can be compensated by increasing the number/rate of packet sent accordingly.

\subsubsection{Stealth vs. Throughput}
We throttle the sender's bandwidth to 100Mb/s-150Mb/s in order not to disrupt the normal network usage on a 1Gb/s LAN. Note that in all cases (except Windows) we could get an almost order of magnitude  improvement in the exfiltration bit rate by consuming almost all the 1Gb/s LAN bandwidth, though this can disrupt the network functionality for other LAN users and end up in exposing the channel. On the other hand, we could throttle the sender and receiver much further, reducing the throughput, but also the probability to get flagged by network IPS/IDS. Specifically, the NetBSD TCP ISN technique only requires a few TCP connections per second, and the Linux technique's sender bandwidth can be reduced by a factor of $n$ if the receiver controls $n$ IP addresses. Additionally, see Appendix~\ref{app:improved-attack-OpenBSD} for stealthier attack variants against macOS and OpenBSD.

\section{Experiments}
\label{sec:experiments}
\subsection{A Setup for the Exfiltration (via DMZ) and Firewall Piercing Experiments}
\label{sec:setup-covert-channel}
The exfiltration experiments have the target host and the sender on 1Gb/s Ethernet network (the {\em lab network} in Israel). \textbf{This network is connected to the Internet via a firewall+NAT device}. Where possible, we ran two tests per attack: one with the receiver in a transcontinental location $A$ (Azure Amsterdam data center), 3300Km from the target host, and one with the receiver in a transatlantic location $B$ (Azure US east coast data center), 9600Km from the target host.
We also measured the RTT average and standard deviation (the latter is indicative of the network jitter) between the two locations and the lab network. The average RTT for $A$ is 79.9ms, with standard deviation 0.3ms, and the average RTT for $B$ is 146.9ms with standard deviation 1.4ms. This is slightly better than the results reported in 2014 \cite{jitter} which is expected in light of the Internet technology improvement over the last seven years.

A single exfiltration test transmits 128 bits from the sender on the internal network, to the receiver on the Internet. This message size was chosen as a benchmark since it demonstrates that AES-128 keys can be exfiltrated in a reasonable amount of time.

Where needed, we used libpcap to craft arbitrary network packets, particularly for spoofing the source IP address of packets, and for accurate timing of packets and packet bursts.

For HTTP/3 server (over QUIC and UDP), we used Caddy Server with the {\tt experimental\_http3} directive. For SSH server (over TCP), we used OpenSSH, and for HTTP/1.1 server (over TCP) we used Apache Web Server. We used Docker version 19.03.8 for the Linux containers. 

\subsection{A Setup for the Host Alias Resolution Experiments}
Given two services (designated by IP address and port), host alias resolution (and de-NATting) provides a single bit of information -- whether the two services reside on the same host (kernel). To implement this, the attacker runs a sender logic for bit ``1'' against one service, and a receiver logic against the other service. If the receiver logic reads ``1'', then the two services run on the same host. If the receiver logic reads ``0'', then the services run on different hosts.

To demonstrate that our results are not randomly correct, we conducted a series of 10 bit ``1'' transfers, for each of the following configurations:
\begin{itemize}
    \item Target addresses on the same host
    \item (Where applicable) target addresses in different containers on the same host (kernel)
    \item Target addresses in different hosts
\end{itemize}

The expected result for the first two configurations, where the addresses are served by the same kernel, is an all-``1'' readout. The last configuration is expected to result in an all-``0'' readout, since the sender changes the state of one kernel, but the receiver probes another kernel. In other words, assuming correct results in the above tests, we are able to tell apart a situation in which the two target addresses are served from the same host (kernel), and a situation in which they are served from two different hosts. This demonstrates host alias resolution, including across containers, as well as de-NATting and container co-residency detection. 

We could only test host alias resolution with Linux, Windows and OpenBSD, because we only had one instance of macOS and NetBSD in our lab. We tested Linux and Windows with an attacker in the lab, because the test involves the attacker spoofing packet source IP addresses, and we categorically do not run such tests over the Internet. With OpenBSD, there is no need to spoof packet source IP addresses, and thus it was tested with an attacker in location $A$.

Other aspects or the experiment are identical to the ones described in Section~\ref{sec:setup-covert-channel}.

\subsection{Description of the Hosts and the Network}

\begin{table*}
\caption{Hosts}
\label{tab:hosts}
\begin{center}
\resizebox{1.0\textwidth}{!}{
\begin{tabular}{|l|l|l|l|l|}
\hline
OS & Hardware & Protocol & Service & Server SW \\ \hline
\begin{tabular}[c]{@{}l@{}}Ubuntu 20.04 \textbf{Linux}\\ (Linux kernel  5.8.12)\end{tabular} & \begin{tabular}[c]{@{}l@{}}Dell E7450: Intel i7-5600U CPU, \\ 16GB RAM, 1Gb/s Ethernet (USB)\end{tabular} & UDP/IPv4 & HTTP/3 & Caddy 2.2.0 \\ \hline
\multirow{2}{*}{\begin{tabular}[c]{@{}l@{}}\textbf{Windows} 20H2\\ build 9042.685\end{tabular}} & \multirow{2}{*}{Intel i7-3770 CPU, 16GB RAM, 1Gb/s Ethernet} & UDP/IPv4 & HTTP/3 & Caddy 2.2.0 \\ \cline{3-5} 
 &  & TCP/IPv4 & HTTP/1.x & Apache 2.2.25 \\ \hline
\begin{tabular}[c]{@{}l@{}}\textbf{macOS} Big Sur 11.1\\ (xnu-7195.60.75-1)\end{tabular} & \begin{tabular}[c]{@{}l@{}}MacBook Air Model A1466: Intel i5-5250U CPU, \\ 8GB RAM, 1Gb/s Ethernet (Thunderbolt)\end{tabular} & UDP/IPv4 & HTTP/3 & Caddy 2.2.0 \\ \hline
\multirow{2}{*}{\textbf{OpenBSD} 6.8} & \multirow{2}{*}{\begin{tabular}[c]{@{}l@{}}Dell E7450: Intel i7-5600U CPU, \\ 16GB RAM, 1Gb/s Ethernet (USB)\end{tabular}} & UDP/IPv4 & HTTP/3 & Caddy 2.2.0 \\ \cline{3-5} 
 &  & TCP/IPv4 & SSH & OpenSSH 8.4 \\ \hline
\multirow{2}{*}{\textbf{NetBSD} 9.1} & \multirow{2}{*}{\begin{tabular}[c]{@{}l@{}}Intel NUC7CJYH: Intel Celeron J4005 CPU, \\ 8GB RAM, 1Gb/s Ethernet\end{tabular}} & TCP/IPv4 & SSH & OpenSSH 8.0 \\ \cline{3-5} 
 &  & TCP/IPv6 & SSH & OpenSSH 8.0 \\ \hline
\end{tabular}
}
\end{center}
\end{table*}

The target hosts used in the experiments are listed in Table~\ref{tab:hosts}. All hosts are located in our lab network, which is a 1Gb/s Ethernet network connected to the Internet over IPv4, through a firewall + NAT device.

\subsection{Experiment Descriptions}
Our results are summarized in Table~\ref{tab:experiments}. Only the sender and receiver locations are listed -- the target host is always on the lab net (except for OF-1). Column ``Section'' refers to the below sections for further information. An empty cell indicates the results are self explanatory.

\begin{table*}
\caption{Experiments}
\label{tab:experiments}
\begin{center}
\resizebox{1.0\textwidth}{!}{
\begin{tabular}{|l|l|l|l|l|l|l|l|l|l|}
\hline
Test & OS & Test Scenario & Protocol & Sender & Receiver & Bits/Repetitions & Success & Bit rate & Section \\ \hline
LE-1 & \multirow{7}{*}{Linux} & Exfiltration (via DMZ) & UDP/IPv4 & Lab Net & Loc. A & 128 & 128/128 & 400 b/h & \ref{sec:LE-1,LE2} \\ \cline{1-1} \cline{3-10} 
LE-2 &  & Exfiltration (via DMZ) & UDP/IPv4 & Lab Net & Loc. B & 128 & 128/128 & 400 b/h & \ref{sec:LE-1,LE2}  \\ \cline{1-1} \cline{3-10} 
LE-3 &  & Exfiltration (container) & UDP/IPv4 & Lab Net & Loc. A & 128 & 128/128 & 400 b/h & \ref{sec:LE-3,LE4} \\ \cline{1-1} \cline{3-10} 
LE-4 &  & Exfiltration (container) & UDP/IPv4 & Lab Net & Loc. B & 128 & 128/128 & 400 b/h & \ref{sec:LE-3,LE4} \\ \cline{1-1} \cline{3-10} 
LF-1 &  & Firewall Piercing (IPfire, slow) & ICMP/IPv4 & Lab Net & Lab Net & 128 & 128/128 & 400 b/h & \ref{sec:LF-1,LF-2} \\ \cline{1-1} \cline{3-10} 
LF-2 &  & Firewall Piercing & ICMP/IPv4 & Lab Net & Lab Net & 128 & 128/128 & 3000 b/h & \ref{sec:LF-1,LF-2} \\ \cline{1-1} \cline{3-10} 
LA-1a &  & Alias Resolution (1 host) & UDP/IPv4 & Lab Net & Lab Net & 10 & 10/10 & N/A & \\ \cline{1-1} \cline{3-10} 
LA-1b &  & Alias Resolution (2 containers) & UDP/IPv4 & Lab Net & Lab Net & 10 & 10/10 & N/A & \\ \cline{1-1} \cline{3-10} 
LA-1c &  & Alias Resolution (2 hosts) & UDP/IPv4 & Lab Net & Lab Net & 10 & 10/10 & N/A & \\ \hline
WE-1 & \multirow{6}{*}{Windows} & Exfiltration (via DMZ) & UDP/IPv4 & Lab Net & Loc. A & 128 & 128/128 & 47.4 b/h & \ref{sec:experiment-windows-server-covert-channel} \\ \cline{1-1} \cline{3-10} 
WE-2 &  & Exfiltration (via DMZ) & UDP/IPv4 & Lab Net & Loc. B & 128 & 128/128 & 47.4 b/h & \ref{sec:experiment-windows-server-covert-channel} \\ \cline{1-1} \cline{3-10} 
WE-3 &  & Exfiltration (via DMZ) & TCP/IPv4 & Lab Net & Loc. A & 128 & 128/128 & 41.4 b/h & \ref{sec:windows-exfil-TCP} \\ \cline{1-1} \cline{3-10} 
WE-4 &  & Exfiltration (via DMZ) & \begin{tabular}[c]{@{}l@{}}UDP/IPv4 (Sender), \\ TCP/IPv4 (Receiver)\end{tabular} & Lab Net & Loc. A & 128 & 128/128 & 47.4 b/h &  \\ \cline{1-1} \cline{3-10} 
WA-1a &  & Alias Resolution (1 host) & UDP/IPv4 & Lab Net & Lab Net & 10 & 10/10 & N/A & \\ \cline{1-1} \cline{3-10} 
WA-1b &  & Alias Resolution (2 hosts) & UDP/IPv4 & Lab Net & Lab Net & 10 & 10/10 & N/A & \\ \hline
ME-1 & \multirow{2}{*}{macOS} & Exfiltration (via DMZ) & UDP/IPv4 & Lab Net & Loc. A & 128 & 128/128 & 1800 b/h & \ref{sec:ME-1,ME-2}\\ \cline{1-1} \cline{3-10} 
ME-2 &  & Exfiltration (via DMZ) & UDP/IPv4 & Lab Net & Loc. B & 128 & 128/128 & 1800 b/h & \ref{sec:ME-1,ME-2} \\ \hline
OE-1 & \multirow{6}{*}{OpenBSD} & Exfiltration (via DMZ) & UDP/IPv4 & Lab Net & Loc. A & 128 & 128/128 & 720 b/h & \ref{sec:experiment-OpenBSD-covert-channel}\\ \cline{1-1} \cline{3-10} 
OE-2 &  & Exfiltration (via DMZ) & UDP/IPv4 & Lab Net & Loc. B & 128 & 128/128 & 720 b/h & \ref{sec:experiment-OpenBSD-covert-channel} \\ \cline{1-1} \cline{3-10} 
OE-3 &  & Exfiltration (via DMZ) & TCP/IPv4 & Lab Net & Loc. A & 128 & 128/128 & 360 b/h & \ref{sec:OE-3} \\ \cline{1-1} \cline{3-10} 
OE-4 &  & Exfiltration (via DMZ) & \begin{tabular}[c]{@{}l@{}}UDP/IPv4 (Sender), \\ TCP/IPv4 (Receiver)\end{tabular} & Lab Net & Loc. A & 128 & 128/128 & 720 b/h & \\ \cline{1-1} \cline{3-10} 
OF-1 &  & Firewall Piercing (slow) & ICMP/IPv4 & Loc. B & Lab Net & 128 & 128/128 & 720 b/h & \ref{sec:OF-1,OF-2,OF-3} \\ \cline{1-1} \cline{3-10} 
OF-2 &  & Firewall Piercing (Esdenera ruleset, slow) & ICMP/IPv4 & Lab Net & Lab Net & 128 & 128/128 & 1200 b/h & \ref{sec:OF-1,OF-2,OF-3} \\ \cline{1-1} \cline{3-10} 
OF-3 &  & Firewall Piercing & ICMP/IPv4 & Lab Net & Lab Net & 128 & 128/128 & 4200 b/h & \ref{sec:OF-1,OF-2,OF-3} \\ \cline{1-1} \cline{3-10} 
OA-1a &  & Alias Resolution 1 host & UDP/IPv4 & Loc. A & Loc. A & 10 & 10/10 & N/A & \\ \cline{1-1} \cline{3-10} 
OA-1b &  & Alias Resolution 2 hosts & UDP/IPv4 & Loc. A & Loc. A & 10 & 10/10 & N/A & \\ \hline
NE-1 & \multirow{4}{*}{NetBSD} & Exfiltration (via DMZ) & TCP/IPv4 & Lab Net & Loc. A & 128 & 128/128 & 7200 b/h & \ref{sec:NE-1,NE-2} \\ \cline{1-1} \cline{3-10} 
NE-2 &  & Exfiltration (via DMZ) & TCP/IPv4 & Lab Net & Loc. B & 128 & 128/128 & 7200 b/h & \ref{sec:NE-1,NE-2} \\ \cline{1-1} \cline{3-10} 
NE-3 &  & Exfiltration (via DMZ) & TCP/IPv6 & Lab Net & Lab Net & 128 & 128/128 & 7200 b/h & \ref{sec:NE-3} \\ \cline{1-1} \cline{3-10} 
NE-4 &  & Exfiltration (via DMZ) & \begin{tabular}[c]{@{}l@{}}TCP/IPv6 (Sender), \\ TCP/IPv4 (Receiver)\end{tabular} & Lab Net & Loc. A & 128 & 128/128 & 7200 b/h & \\ \hline
\end{tabular}
}
\end{center}
\end{table*}

\subsubsection{Linux Exfiltration (via DMZ) over UDP/IPv4}
\label{sec:LE-1,LE2}
In tests LE-1 and LE-2 (locations $A$ and $B$, resp.), the target is an Intel x64 machine, therefore $f$=250Hz. The covert channel was defined with $\Delta t=10ms$ (considerably lower values are not usable due to the jitter). Theoretically, we could have used $M=3$ with this $\Delta t$, but we set $M=6$ to eliminate the occasional timing deviation in sampling caused by jitter. 
Due to logistic and budgetary constraints, the receiver in our experiments used only a single IP address.

During the preliminary tests, we noticed that the target machine's CPU was mildly loaded during the transmissions of the bit ``1''. This is not expected to happen in a production system, where the server hardware is typically much more powerful than our Dell machine. However, given the Dell machine's CPU load, one could argue that the effects we observe have to do with CPU load (either due to application processing, or due to network processing). Ironically, this points at another potential covert channel -- measuring the effects of CPU load on the network traffic shape. But as we mention above, this is unlikely to happen in a production network, where the CPU power is more in balance with the network capacity.

To make sure the effect we measure is only related to the IPv4 ID information leakage, we altered the channel definition for the purpose of the experiment. Instead of staying idle (not transmitting anything) for bit ``0'', in the experiment, we transmit the same amount of packets, but using a small subset of 90 spoofed source IP addresses whose counters do not collide with the receiver's counter. A-prioi, choosing 90 counters out of 2048 yields a good likelihood of this condition to be met. A-posteriori, we can verify this easily. If the receiver's IP address IP ID counter does collide with one of the 90 counters, the vector of bits the receiver reads will be all 1's. We verified this is not the case. This revised experiment ensures that the host's CPU load is comparable for both ``0'' and ``1'' bits.

Given the above results, we estimate that the bit error rate in this channel is below 0.004. In this case, simple error correction code can be used to correct any sparse errors.

\subsubsection{Linux Exfiltration via Containers over UDP/IPv4}
\label{sec:LE-3,LE4}
The LE-3 and LE-4 tests (locations $A$ and $B$ resp.) are very similar to the previous section, with two co-resident containers running on a single host (kernel). The sender resides in the internal (isolated) network, communicating with the internal (isolated) container over HTTP/3, and the receiver resides on the Internet, communicating with the second container over HTTP/3. In all other aspects, these tests are identical to the one in the previous section.

\subsubsection{Linux Firewall Piercing over ICMP/IPv4}
\label{sec:LF-1,LF-2}
Due to technical reasons, we could only conduct the Linux firewall piercing attacks in the lab. For test LF-1, we installed a typical Linux-based firewall image (IPfire 2.25 (x86\_64) running Linux kernel 4.14.112) on a dedicated device, and configured it to block communications between the internal ``green'' network and the external ``red'' network (an Internet-facing lab network). We then ran our attack, but for simplicity of testing, we sent data from the red network to the green network. The data transfer rate was limited by the sender machine's effective bandwidth -- we could only send up to $\approx$14,000 ICMP packets/s. A stronger sender machine would have achieved a higher throughput.

In order to compare the speed of ICMP exfiltration to that of QUIC/UDP exfiltration, we conducted another experiment (LF-2), this time using the same target machine as in \ref{sec:LE-1,LE2}. We used a stronger sender that was able to send $\approx$130,000 ICMP packets/s, and the throughput was 1.2 bits/s (3000 b/h), which is 7.5 times faster than the QUIC/UDP variant. Here too, we could use a stronger sender to get even higher exfiltration throughput.

\subsubsection{Windows Server Exfiltration (via DMZ) over UDP/IPv4}
\label{sec:experiment-windows-server-covert-channel}
We did not have at our disposal a Windows Server on a network that allows spoofing, so instead, for tests WE-1 and WE-2 (locations $A$ and $B$, resp.), we simulated Windows Server on a Windows 10 platform. This was done by setting {\tt TcpipIsServerSKU} to 1 and the default compartment's IPv4 {\tt PathSet} purge threshold (32 bit quantity at offset 332 of {\tt PathSet}) to 32768. This made the system behave identically to Windows Server for the purpose of our experiments.

For the channel definition, we used $K=6$. A single bit transfer took 76s (theoretically this could be done with 71s per bit, but due to technical issues, and to make sure the channel can handle some organic network traffic, we padded it by 5s).

\subsubsection{Windows Exfiltration (via DMZ) over TCP/IPv4}
\label{sec:windows-exfil-TCP}
In experiment WE-3, we demonstrated that the Windows covert channel can be used over TCP. In essence, this is a very similar experiment to the one described in Section~\ref{sec:experiment-windows-server-covert-channel},  but this time we used TCP. 
The sender sent SYN packets with spoofed source IP address to the target host. These packets are short (40 bytes at the IPv4 level) 
 and can be transmitted rapidly. 
Assuming the spoofed IP addresses are not associated with any machines, there will not be any transmission from these IP addresses following the spoofed SYN packets. Windows responds immediately with a SYN+ACK packet, with additional two retransmissions (the default ``Max SYN Retransmissions'' is 2) at exponential backoff, so the first retransmission is sent after an ``Initial Retransmission Time-Out'' seconds, henceforth $\RTO$ (in Windows Server, by default $\RTO$=3s), and the second retransmission packet is sent $2\cdot \RTO$ after the first retransmission. These packets are not answered of course, so finally Windows sends an RST packet, $4\cdot \RTO$ seconds after the third SYN+ACK packet is sent ($7 \cdot \RTO$ seconds after the SYN packet is received). Therefore, Windows sends the last packet to a spoofed IP address 21s after a SYN is received from it. This overlaps with the 10 seconds the receiver needs to wait after the sender's burst in order for the purge to be concluded, so only 11 seconds are actually added to the time-per-bit compared to the UDP case. In a lab experiment, we found that Windows' SYN cache is not exhausted even by 13000 half-open connections, thus our technique for TCP/IPv4 in Windows is non-disruptive.

As for the receiver, we implement it to send an RST packet to the target host as soon as the target host's SYN+ACK arrives at the receiver, thus the {\tt Path} object's last access time is only extended by the RTT between the target host and the receiver, having a negligible impact on the attack timing.

We carried out the experiment similar to Section~\ref{sec:experiment-windows-server-covert-channel}, with the modified attack timing. Since we used Windows 10 as a target host, we had to adjust its {\tt Initial RTO} value to 3000ms to simulate Windows Server (the default Windows 10 value is 300ms). 
\subsubsection{macOS Exfiltration (via DMZ) over UDP/IPv4}
\label{sec:ME-1,ME-2}
In tests ME-1 and ME-2 (locations $A$ and $B$, resp.), the channel was defined with $K=700$, thus ensuring that the probability for a sent bit ``1'' to be received as ``0'' is very low ($\nicefrac{1}{1767}$). The sender transmitted 5000 packets per bit. Theoretically, the sender needed to send 4096 packets to signal the bit ``1'', but in practice, due to server responsiveness issues, we had the sender send some extra packets. With throttling, this took 0.39 seconds. Due to some overheads incurred by our specific test implementation, each bit transmission took two seconds. 

\subsubsection{OpenBSD Exfiltration (via DMZ) over UDP/IPv4}
\label{sec:experiment-OpenBSD-covert-channel}
In tests OE-1 and OE-2 (locations $A$ and $B$, resp.), the channel was defined with $K=700$, thus ensuring that the probability for a sent bit ``1'' to be received as ``0'' is very low ($\nicefrac{1}{1767}$). The sender transmitted 40000 packets per bit. Theoretically, the sender needed to send 32768 packets to signal the bit ``1'', but in practice, due to server responsiveness issues issues, we had the sender send 40000 packets. This took 3.32 seconds. Due to some overheads incurred by our specific test, each bit transmission took 5 seconds. 

\subsubsection{OpenBSD Exfiltration (via DMZ) over TCP/IPv4}
\label{sec:OE-3}
In test OE-3, the sender sent SYN packets to the target host. 
Both the sender and the receiver were implemented to send an RST packet to the target host as soon as the target host's SYN+ACK arrives at the sender/receiver, to ensure no retransmissions clutter the network and that the host's SYN cache is not overflowed, since the pending connection is removed from it once the RST packet for it arrives.

We throttled the sender to about 6000 packets/sec in order to make sure the host's TCP SYN cache is not exhausted (the OpenBSD SYN cache size limit is 10255 entries). Due to the throttling, the time per bit was extended to 10s.

\subsubsection{OpenBSD Firewall Piercing}
\label{sec:OF-1,OF-2,OF-3}
Test OF-1 was conducted over the Internet, with a very weak machine running OpenBSD 6.7 in Location $B$. That machine could only handle $\approx$9,000 ICMP Echo Requests per second, which set the constraint on the throughput. We achieved a single bit transfer in 5s. 

We also ran in-the-lab tests with firewalls (we could not deploy Internet-facing firewalls). In test OF-2, we simulated an OpenBSD-based firewall. For this, we used a PF ruleset we extracted from an Esdenera Firewall-3 installation \cite{esdenera}. This ruleset explicitly allows ICMP traffic to and from the firewall, on all interfaces. We tested OpenBSD 6.9 in this case.
In test OF-3, we measured the speed of the attack with a sender capable of sustaining $\approx$68,000 ICMP Echo Requests per second, against the usual Dell target running OpenBSD 6.8, which can serve $\approx$260,000 Echo Requests per second. We obtained a 0.85s bit transfer time, but using a stronger sender machine can significantly increase this figure.

\subsubsection{~~NetBSD Exfiltration (via DMZ) over TCP/IPv4}
\label{sec:NE-1,NE-2}
Our covert channel for NetBSD over IPv4 is not in itself interesting, because NetBSD already has a TCP/IPv4 covert channel in the form of its global IPv4 ID counter. However, we wanted to emphasize the applicability of our TCP/IPv4 technique over the Internet since we cannot demonstrate our TCP/IPv6 technique over the Internet due to technical constraints. By demonstrating our TCP/IPv4 covert channel over the Internet, we infer that the same holds for our TCP/IPv6 channel.

In our NE-1 and NE-2 experiments (with senders at $A$ and $B$, resp.), we set $\Delta t$=0.2s, with total time per bit 0.5s (this is due to some limitations of our testing code, and we note that theoretically NetBSD's covert channel can be run at much higher speeds), thus theoretically threshold for $m-m'$ should have been 2, but we set it to 3 to avoid false positives from occasional TCP connections established with the target host. 

\subsubsection{~~NetBSD Exfiltration (via DMZ) over TCP/IPv6}
\label{sec:NE-3}
In test NE-3 we had to use a local receiver since our lab does not have IPv6 connectivity to the Internet. 
\section{Remediation and Suggestions}
\label{sec:remediation}
We begin with the easy parts, in which the attack can be completely eliminated:
\begin{itemize}
    \item TCP/IPv4 packets, and IPv4 packets with DF=1 -- these packets can have ID=0 or a completely random ID. In Linux, all TCP packets, and UDP packets whose length is below the PMTU are sent with DF=1. In Windows and macOS, all TCP packets are sent with DF=1. Moreover, in macOS, the TCP packets have ID=0, which demonstrates that simply setting ID=0 suffices for these cases.
    Note that e.g. QUIC {\em requires} setting DF=1 in UDP \cite[Section 14]{rfc9000}, but Caddy Server leaves the choice of DF to the operating system.
    \item TCP ISN -- this field should not contain state information, i.e. should contain random data and client information only.
    \item IPv6 Flow label (see Appendix~\ref{app:NetBSD-flowlabel-cryptanalysis} and Appendix~\ref{app:NetBSD-flowlabel-bit-flipping}) -- can be chosen at random (i.e. using a ``regular'' PRNG). There is no RFC requirement for the flow label to have a unique value in any time duration.
\end{itemize}
In some operating systems, a built-in filtering functionality facilitates overriding a protocol field value with a fixed/random value. 
See Appendix~\ref{app:nftables} for Linux nftables examples.

The challenging case is UDP/IPv4 (and ICPMv4/IPv4) ID with DF=0. In this case, there is a real need for the ID field to be non-repeating to some extent. We can make the following suggestions to reduce the attack surface and eliminate some attack variants, but the underlying issue is still in effect even with our suggestions:
\begin{itemize}
    \item Maintain a separate IPv4 ID state according to the destination address: one state for RFC1918 destination addresses (internal networks) and another state for external destinations. This eliminates the exfiltration attacks where the sender is on an internal network and can only spoof internal network addresses. 
    \item Maintain an IPv4 ID generator state per service (i.e. per a listening IP address and port). This eliminates the host alias resolution attacks. A weaker version is to maintain a state per listening IP address.
    \item Maintain an IPv4 ID generator state per container -- this eliminates the cross-container attacks.
\end{itemize}
Another approach is to ensure that connection-less transport protocols emit IPv4 packets whose length is below a reasonably chosen minimal PMTU, and simply set DF=1 and ID=0 for {\em all} these packets. For example, DNS over UDP/IPv4 can function with message length limit of 512 bytes (540 bytes including UDP and IPv4 headers), QUIC over UDP/IPv4 can function with message length limit of 1200 bytes (1228 bytes including UDP and IPv4 headers), and ICMP Echo Requests whose size exceeds PMTU can be discarded.

Finally, it is possible to simply randomize the IP ID, which guarantees maximum security, at the expanse of somewhat increased malformed/dropped packet rate (due to the possible ID collisions and their effect on fragment reassembly). 

Note that ICMP rate limits do not eliminate the covert channels: even an aggressive per-destination rate limit is ineffective against the attack, as the attacker may use multiple, possible spoofed, source IP addresses. A global rate limit, on the other hand, facilitates a covert channel via overflowing (or not) the packet counter.

\section{Conclusions}
\label{sec:conclusions}
We conclude that global states in network protocols of server hosts can be exploited in various ways to weaken the security of networks where such hosts reside, and to glean information on the implementation of services which is otherwise abstracted from their consumers.

While a global state can be as obvious as an incremental counter, we also demonstrated that it can take a more subtle form. For example in the case of OpenBSD and macOS IPv4 ID, the observation of collisions in IPv4 ID values (or lack thereof) is an indication of the internal global state, which is not directly exposed to the attacker.
Other such non trivial examples are the Linux IPv4 ID generation scheme, and the Windows IPv4 ID generation scheme. 

Our work focused on {\em firewalled} hosts, which is a very common scenario for Internet-facing servers. A stateful firewall eliminates many known information leakage attacks, and therefore poses a serious obstacle for techniques that rely on information leakage via a global protocol state. We described three new attack scenarios based on information leakage via a global protocol state: firewall piercing, exfiltration form a secure network, and exfiltration across containers. We demonstrated our three new scenarios, as well as a host alias resolution scenario and a host alias resolution across containers scenario. All these scenarios were demonstrated with the sender (and the host, where applicable) behind a firewall, in realistic conditions, and over the Internet where possible. We also demonstrated some cross protocol attacks, e.g. where the sender uses UDP/IPv4 and the receiver uses TCP/IPv4, and where the sender uses TCP/IPv6 and the receiver uses TCP/IPv4. We measured the exfiltration bandwidth achieved, and found it to be sufficient for e.g. key material exfiltration in hours or minutes.

The problem of {\em secure} IPv4 ID generation seems to be theoretically intractable, due to the conflicting requirements -- generating long sequences of unique values, at least per $(\IP_{\src},\IP_{\dst})$ and leaking no information, while maintaining a ``reasonably sized'' state, which is much smaller than the number of expected concurrent tuples. Intuitively, this is due to the pigeonhole principle, as it is impossible to maintain an independent state per tuple with less bits than the number of concurrent tuples. However, we do list some recommendations that considerably reduce the attack surface and specifically address the attacks scenarios we described.

It is important to understand that information leakage from network protocol global states can affect the security of networks and systems in surprising and unexpected ways. In our research, we highlight three novel attack scenarios. 
It is quite possible that additional attack scenarios exist, and thus there is room for future research on this topic.

\section{Vendor Status}
\label{sec:vendor-status}
All vendors were notified on January \nth{27}, 2021. 
\begin{itemize}
\item \textbf{Linux} issued a patch \cite{Linux-patch} for kernel version 5.13-rc1 that significantly increases the size of the $\beta$ table. This patch was back-ported to versions  5.12.4, 5.11.21, LTS branches 5.10.37, 5.4.119, 4.19.196, 4.14.238, 4.9.274 and 4.4.274. Back-ports to all 4.x branches were provided by us based on the back-port to the 5.x versions. This issue is tracked as CVE-2021-45486. 

Linux also issued a patch replacing the existing IPv6 ID algorithm (similar to the IPv4 ID algorithm) with a completely randomized IPv6 ID \cite{Linux-IPv6-ID-patch} which is included in version 5.14-rc1, and was back-ported to 
kernel versions 5.13.3, 5.12.18, 5.10.51, 5.4.133, 4.19.198, 4.14.240, 4.9.276 and 4.4.276. This issue is tracked as CVE-2021-45485.

\item \textbf{Microsoft} informed us on March \nth{15}, 2021, that ``[Microsoft] determined your finding does not meet our bar for servicing''.

\item \textbf{Apple} released a fix for the IPv4 ID issue in macOS 12.1 \cite{apple-2021-patch-mac-monterey}, macOS 11.6.2 \cite{apple-2021-patch-mac-bigsur}, macOS 10.15 \cite{apple-2021-patch-mac-catalina}, iOS/iPadOS 15.2 \cite{apple-2021-patch-ios} tvOS 15.2 \cite{apple-2021-patch-tvos} and watchOS 8.3 \cite{apple-2021-patch-watchos}. 

The ICMP rate limit issue was reported to Apple separately on June \nth{11}, 2021, and regarding it, Apple informed us on July \nth{21}, 2021, that ``while we [Apple] do not see any security implications, we have forwarded it on to the appropriate team to investigate for potential future enhancements''. \item \textbf{OpenBSD} has not responded to multiple requests for status, has not informed us about any decision/resolution, and has not fixed the issue in their public source code repository.

\item \textbf{NetBSD} produced a patch \cite{NetBSD-patch} for NetBSD 9.2 that better randomizes the TCP ISN (CVE-2021-45488), IPv6 Flow Label (CVE-2021-45489) and IPv6 ID (CVE-2021-45484). This patch also switches the NetBSD implementation of the IPv4 ID field from a simple counter to a random number generator (CVE-2021-45487).
\end{itemize}

\section*{Acknowledgment}
We would like to thank Benny Pinkas, the anonymous NDSS 2022 reviewers and Haixin Duan (our NDSS shepherd) for their valuable feedback, and Ehood Porat for his help in reverse engineering some parts of Windows code.

\bibliographystyle{plain}

\appendix

\section{Experiment: A Survey of Responses from Widely-Used Internet Servers to UDP, TCP and ICMP Probing}
\label{app:probing}

We estimate the portion of widely-used Internet hosts which do not respond to a UDP packet to a closed port, i.e. hosts that do not send back an ICMP message of any type (or any non-ICMP message) for such invalid packets. We also estimate the portion of widely-used Internet hosts which silently drop TCP SYN packets to closed TCP ports. Finally, we estimate the portion of widely-used Internet hosts that respond to ICMP Echo Request (via the Ping utility).

We use the top 1000 domains in the Alexa top 1M domain list obtained on December \nth{13}, 2020 as a starting point. We then query for the DNS NS (name server) record-set for each domain, take the first NS record from the answer, and resolve its IPv4 address (by sending a DNS A query and take the first record from the answer). This yields an IP address of an authoritative DNS server for the domain at hand. This host answers UDP queries to port 53 (DNS). As such, it serves as a model for a reliable host that runs a UDP-based service, similar to a QUIC server. Moreover, since the host is an authoritative DNS server for a popular domain, we expect it to be highly available, i.e. we can assume there are no downtime/service issues for this host. We eliminated the duplicates in the IP address list, and ended up in a list of 748 unique IP addresses. 

\textbf{UDP Closed Ports:} For each IP address, we sent a UDP packet for three randomly chosen high ports. These UDP ports are un-assigned by IANA, and have no popular service associated with them, hence are very likely to be closed. We observe whether we receive any response from the host, either over UDP or over ICMP. As expected, the only responses we received were ICMP Type 3 messages (``Destination Unreachable''), mostly with Code 3 (``Port Unreachable''), but in three cases, we received Code 10 (``Communication with Destination Host is Administratively Prohibited'') and in two cases we received Code 13 (``Communication Administratively Prohibited''). 

\textbf{TCP Closed Ports:} We sent three TCP SYN packets to each port used above, and recorded whether we received any response back. As expected, most servers that responded, did so with an RST packet, although one server responded with SYN+ACK on two ports, which is likely a deceptive practice. Regardless, we count this server as responsive to TCP probes for closed ports.

\textbf{ICMP Echo Requests:} We sent each server three ICMP Echo Request messages, and we recorded whether we received any ICMP Echo Reply messages back.

The results are as follows: 116 (15.5\%) servers responded to the UDP probing, and 59 (7.9\%) servers responded to the TCP probing. The overlap was significant -- 42 servers responded to both UDP and TCP probing. Overall, 133 (17.8\%) servers responded to either UDP or TCP.
667 (89.2\%) servers replied to ICMP Echo Requests. This set almost completely covers the set of servers that responded to either UDP or TCP probing -- the overlap was 125 servers. In other words, the ICMP probing is almost a complete superset of the TCP or UDP probing -- it misses about 1\% of the space covered by UDP or TCP, and contributes additional 72.5\% of coverage.

\subsection{Improved Attacks}
\subsubsection{Improved UDP/IPv4 ID Attack for Linux}
\label{app:improved-attack-Linux}
An improvement to the technique described in Section~\ref{sec:linux-exfil} is as follows: assume there also exist a one-way channel from the receiver to the sender (which can be implemented along the same lines of the sender-to-receiver channel). In such a case, the improved technique works in two phases. In the first phase, the sender singles out one spoofed IP address which hashes to the same counter with the receiver. For this phase, a feedback is needed from the receiver. In the second phase, the sender needs to spoof $2M-1$ packets from a single address in order to transfer a single bit ($L$ times more efficient in packets than the original technique), and the receiver can read one bit per $\Delta t$ time slot, which yields a $\frac{(2M-1) L}{B\Delta t}$ faster transfer rate. In the first phase, the sender and receiver engage in a binary search for a spoofed address that hashes to the receiver's counter. This process involves sending $(2M-1)L$ packets in $\lceil \log_2 L \rceil$ iterations. The sender can assume, by choosing large enough $L$, that at least one spoofed address will have a counter collision in the target host with the receiver's IP address, and the first phase aims to find one such spoofed address. In the first iteration, the sender picks $\frac{L}{2}$ spoofed addresses and sends the bit ``1''. The receiver sends back the bit it read. If it is ``1'', the sender deduces that there is colliding spoofed address in the sub-list, and proceeds with the binary search on this list. If the bit is ``0'', the sender deduces that there is a colliding spoofed address in the complementary sub-list, and proceeds with the binary search on the complementary sub-list. At the end of this phase, the sender has one address which hashes to the same counter used for the receiver's address. The second phase can then proceed using a single address. Note that the second phase does not need the receiver-to-sender channel.

\subsubsection{A Stealthier IPv4 ID Attack for macOS and OpenBSD}
\label{app:improved-attack-OpenBSD}
For macOS and OpenBSD, the IPv4 ID attack can be made stealthier by having the sender and receiver run ``native'' QUIC sessions with the target host, for example HTTP/3 resource downloads. For example, the sender can request a 4.9152MB (macOS) or 39.3216MB (OpenBSD) file (or larger) from the target host. This ensures that at least 4096 (macOS) or 32768 (OpenBSD) QUIC packets are sent from the target host to the sender (as each QUIC packet carries at most 1200 payload bytes). Likewise the receiver can request a 840KB file (or larger) file from the target host to force the host to send the receiver at least 700 QUIC packets. Of course, instead of requesting one file (or resource), the sender and the receiver can request multiple resources so that the total number of packets sent by the host is above the respective thresholds.

This can also be applied to OpenBSD with TCP traffic. In this case, there is an additional benefit, which is that there is no pressure on the TCP SYN cache in the target host.

\subsection{Additional Information Leakage Attacks}
\label{app:more-attacks}

\subsubsection{A Covert Channel via Overflowing the NetBSD and OpenBSD TCP SYN Cache}
\label{app:SYN-Cache-Overflow}
NetBSD's TCP SYN cache is a hash table containing 293 buckets with up to 105 entries each, and with a global limit of 10255 entries. An inbound SYN packet is hashed into a bucket based on the source IP address, the source port, the destination port and a secret 64-bit key. Then a SYN+ACK is sent, and the bucket entry becomes a half-open connection. Statistically, for evenly distributed traffic, the global limit is reached before any bucket becomes full. When the global limit is reached, the half-open connection's bucket\footnote{The bucket from which the entry is removed is the first non-empty bucket in the list of buckets, starting from the half-open connection's bucket. When the traffic is evenly distributed, this will be the half-open connection's bucket.} has its oldest entry removed to keep the hash table from overflowing. NetBSD responds to TCP SYN packets as follows: a first SYN+ACK packet is sent immediately. A second SYN+ACK is sent after three seconds and three more SYN+ACK packets are sent with exponential backoff factor of 2. This facilitates a side channel: let us assume that the target NetBSD host has it SYN cache table relatively empty. An observer can send $K$ TCP SYN packets to an open port on a NetBSD host, and then observe the SYN+ACK packets sent back from the host, without completing the handshake. If, during the 45 second period after the SYN packets were sent, the TCP SYN cache becomes full due to another party sending SYN packets, and some of the old packets' entries get evicted, the host will not send any further SYN+ACK packets for these evicted entries. If there were $F$ SYN packets sent after the SYN cache became full, the observer expects $K(1-\frac{1}{293})^F \approx Ke^{-\frac{F}{293}}$ ``surviving'' half-open connections for which the host continues to send SYN+ACK packets.

We now show how to turn this into a covert channel. The receiver sends $K$ SYN packets from $K$ different TCP source ports. Then the sender can send $(10255+\mu)$ SYN packets from different source ports to the host to signal the bit ``1''. Note that $F=K+\mu$, so the receiver observes that the host continues to send SYN+ACK only for $Ke^{-\frac{K+\mu}{293}}$ half-open connections (in expectation). 
To send the bit ``0'', the sender does nothing, and the receiver observes that for all ports (up to packet loss), the host continues sending SYN+ACK.

For example, setting $K=10, \mu=1000$ yields expected 0.32 open connections (out of 10 established by the receiver) to continue sending SYN+ACK, with $\sigma=0.56$ (clearly this is not normally distributed, but this $\sigma$ value still indicates that it is extremely unlikely to get a value above 5, for example). Thus, the receiver sends 10 SYN packets to the host, the receiver sends (or does not send) 11225 SYN packets to the host, and if the number of half-open connections for which the host continues to send SYN+ACK to the receiver is less than 5, the receiver deems that the sender signalled the bit ``1'', and otherwise, the receiver deems that the sender signalled the bit ``0''.

This attack is quite expensive in terms of packet count compared to the NetBSD TCP ISN attack. Moreover, it is disruptive, since SYN packets from organic clients of the host may get evicted from the SYN cache which results in TCP connection failures.

The OpenBSD SYN cache is almost identical to the NetBSD one, with the following exception: in OpenBSD, when 100000 insertions are made to the current TCP SYN cache, it is frozen, and another SYN cache instance is used as the active SYN cache for new insertions. The impact on our technique is that about once in 9 bits, the transmission fails because the old half-open connections from the receiver are frozen and will not be removed (until they expire naturally).

\subsubsection{A Covert Channel via NetBSD TCP/IPv6 Flow Label -- Cryptanalysis}
\label{app:NetBSD-flowlabel-cryptanalysis}
NetBSD's TCP/IPv6 flow label field (20 bits) is generated by the A2/20 PRNG (following the naming convention introduced in \cite{BSD-IPID-attack}). The flow label field comprises a most significant bit (MSB) part, and 19 least significant bits part. The PRNG is reseeded every 180 seconds or 200000 internal PRNG steps (whichever comes first). As part of the reseeding procedure, the parameters for generating the least significant 19 bits are sampled from a strong kernel PRNG, and the MSB is flipped.
Since the MSB is static between reseedings, it can be ignored for the remaining of the discussion. The algorithm is designed to produce unique outputs during two consecutive reseeding periods. This is not a required property for IPv6 flow label, and is probably a by-product of deriving this algorithm from the A2/16 algorithm used for IPv4 ID generation back in 2003, which does require this uniqueness property.

The least significant 19 bits of the PRNG are generated using Algorithm~\ref{alg:NetBSD-flowlabel}. Note that in this sub-section, $M$ denotes a parameter of Algorithm~\ref{alg:NetBSD-flowlabel}, instead of its role in the main paper. In a single PRNG invocation, the PRNG draws a random number $1 \leq n \leq 4$ and runs $n$ internal steps. NetBSD's flow label generation involves {\em two} invocations of the PRNG, but only the output of the latter one is used for the flow label field. Thus overall, flow label generation consumes 2-8 internal PRNG steps (5 internal steps on average). 

Note that the pseudo-random flow label is only used by a NetBSD server {\em after} the TCP connection is established. The flow label for a NetBSD TCP/IPv6 SYN+ACK packet is 0 (however, RST packets always carry a random flow label generated with Algorithm~\ref{alg:NetBSD-flowlabel} -- and this facilitates idle scanning). Also, while our interest is in NetBSD as a TCP/IPv6 server, our analysis also covers the case of flow labels generated by NetBSD as a TCP/IPv6 client. Finally, note that our analysis can be applied, with very minor modifications, to NetBSD's IPv6 ID generation algorithm (A2/32).

\begin{algorithm}[ht]
\caption{NetBSD IPv6 Flow Label Algorithm (19 Least Significant Bits)}
\label{alg:NetBSD-flowlabel}
\begin{algorithmic}[1]
\State \Comment{$M=279936=2^7 \cdot 3^7$}
\State \Comment{$N=524269$ (prime), $\varphi(N)=N-1=2^2\cdot 3^2 \cdot 14563$}
\Procedure{Reseed}{}
\State $x \gets \Call{random}{\{0,\ldots,M-1\}}$ 
\State $s_1 \gets \Call{random}{\{0,\ldots,2^{19}-1\}}$  
\State $s_2 \gets \Call{random}{\{0,\ldots,N-2\}}$  
\State $a \gets 7^{2 \cdot \Call{random}{\{0,\ldots,2^{19}-1\}}} \mod M$  
\State $b \gets \Call{random}{\{x|0<x<M \land \gcd(x,M)=1\}}$
\State $g \gets 2^{\Call{random}{\{x|0<x<N \land \gcd(x,N-1)=1\}}} \mod N$
\EndProcedure
\Procedure{Generate-FlowLabel}{}
\State $n \gets \Call{random}{\{1,\ldots,4\}}$
\For{$i=1$ to $n$}
\State $x \gets (ax+b) \mod M$
\EndFor
\State return $s_1 \oplus (g^{x+s_2} \mod N)$ 
\EndProcedure
\end{algorithmic}
\end{algorithm}

We now cryptanalyze Algorithm~\ref{alg:NetBSD-flowlabel}. We assume that an attacker collects consecutive NetBSD flow labels, i.e. flow labels generated for consecutive TCP/IPv6 connections. We allow a certain degree of noise, i.e. out-of-order values, and missing values. However, we do not allow values not generated by the algorithm in the input data. Our analysis requires a large portion of {\em pairs} of values to be correct, where a pair is two consecutive flow label values. We do not require {\em all} pairs to be correct though. Our cryptanalysis uses many concepts and techniques from the cryptanalysis of A2/20's sibling algorithms, A0/16 and X3/16, as reported in \cite{BSD-IPID-attack}. However, applying the techniques for the 16 bit algorithms to the 20 bit algorithm results in an unacceptably long (though feasible) run time -- specifically, ``Phase 2'' in \cite{BSD-IPID-attack} involves enumeration over all $(N-1)\frac{\varphi(N-1)}{2}$ combinations of valid $s_2$ and $g$ values, which for A2/20 would yield $2^{35.41}$ values, and then going over all pairs (if some noise is expected). This can take over an hour when run single-threaded on a modern CPU. Our aim in the improved cryptanalysis herein is to provide an algorithm whose speed is 4.5-5 orders of magnitude higher than the techniques presented in \cite{BSD-IPID-attack}. Indeed, in our test, the typical attack runtime, when running single-threaded on a modern CPU, is less than 50ms.

Our cryptanalysis comprises four phases: in the first phase, we use consecutive pairs to extract $s_1$ (with enumeration over $2^{19}$ values). This is very similar to \cite[Section 3]{BSD-IPID-attack}. In the second phase, we extract $g$. In phase 3, we extract $s_2$ in part. In phase 4, we enumerate over the possible $s_2$ values, and we build $a$ and $b$ bit by bit (almost literally). In this phase we also need pairs.

We assume the attacker has $L$ individual flow label values $0 \leq F_i < 2^{19}$ (we ignore the most significant bit of the flow label), and among them the attacker has $P$ pairs of consecutive flow label values $(F_i,F'_i)$. Clearly, $L\ge P+1$, with equality holds when the values are consecutive flow labels.

\textbf{Phase 1 -- Extracting $s_1$:}
Offline, attacker prepares a $\log_2(\cdot)$ table for $\{1,\ldots,N-1\}$, i.e. a table wherein $(2^{n} \mod N) \mapsto (n \mod (N-1))$. 

Online, the attacker enumerates over all possible $2^{19}$ values of $s_1$ and analyzes the pairs $(F_i,F'_i)$. For the correct guess of $s_1$, the following will hold: $\log_2(F_i \oplus s_1)=(\log_2 g)(x_i+s_2) \mod (N-1)$, and likewise for $F'_i$. Subtracting, we get: 
$$\log_2(F'_i \oplus s_1)-\log_2(F_i \oplus s_1) \mod (N-1) = $$
$$(\log_2 g)(x'_i-x_i) \mod (N-1)$$
Taking modulo 12 on both sides, keeping in mind that $12|(N-1)$, we get:
$$\log_2(F'_i \oplus s_1)-\log_2(F_i \oplus s_1) \mod 12 = $$
$$(\log_2 g)(x'_i-x_i) \mod 12$$
Now, since $12|M$, and since $a \mod 12=1$ by construction, we have $x'_i-x_i \mod 12 = n_i \cdot b$ where $n_i$ is the number of internal steps between $x_i$ and $x'_i$ (see lines 10-12 of Algorithm~\ref{alg:NetBSD-flowlabel}), keeping in mind that the algorithm is executed twice per flow label, i.e. $2 \leq n_i \leq 8$. Thus:
$$\log_2(F'_i \oplus s_1)-\log_2(F_i \oplus s_1) \mod 12 = $$
$$((\log_2 g)b)n_i \mod 12$$
Note that since $g$ is a generator in the multiplicative group modulo $N$, $\log_2 g$ is invertible modulo $\varphi(N)=N-1$, as well as modulo 12. $b$ is invertible modulo $M$ by construction, so $(\log_2 g)b$ is invertible modulo 12. Therefore, $((\log_2 g)b \mod 12$ is invertible, and so can only take four values: 1, 5, 7 or 11. Each value induces a set of 7 possible values (only) for $\log_2(F'_i \oplus s_1)-\log_2(F_i \oplus s_1) \mod 12$, conforming to the 7 possible values of $n_i$. If $((\log_2 g)b \mod 12=1$, then $\log_2(F'_i \oplus s_1)-\log_2(F_i \oplus s_1) \mod 12 \in \{2,3,4,5,6,7,8\}$. If $((\log_2 g)b \mod 12=5$, then $\log_2(F'_i \oplus s_1)-\log_2(F_i \oplus s_1) \mod 12 \in \{1,3,4,6,8,10,11\}$, and so forth. 

We see, therefore, that for the correct guess of $s_1$, the $P$ quantities $\log_2(F'_i \oplus s_1)-\log_2(F_i \oplus s_1) \mod 12$ all belong to one of four sets of seven values (out of 12 possible values). Denote by $C_j=|\{i|\log_2(F'_i \oplus s_1)-\log_2(F_i \oplus s_1) \mod 12=j\}|$ -- the number of times the value $j \in \{0,\ldots,11\}$ appears in the list of values $\log_2(F'_i \oplus s_1)-\log_2(F_i \oplus s_1) \mod 12$. Define:
$$X_1=C_2+C_3+C_4+C_5+C_6+C_7+C_8$$
$$X_5=C_1+C_3+C_4+C_6+C_8+C_{10}+C_{11}$$
$$X_7=C_1+C_2+C_4+C_6+C_8+C_9+C_{11}$$
$$X_{11}=C_4+C_5+C_6+C_7+C_8+C_9+C_{10}$$
If all $P$ pairs are {\em true} (i.e. they are indeed pairs of consecutive outputs from the PRNG), we will have $X_1=P$ or $X_5=P$ or $X_7=P$ or $X_{11}=P$. If there are some false pairs among the $P$ pairs, this will not hold as-is, but we can still use a statistical approach. For an incorrect $s_1$, a single counter $C\sim\mathit{N}(\frac{P}{12}, \frac{11}{144}P)$, and a sum of 7 counters has distribution $\mathit{N}(\frac{7}{12}P, 7\cdot\frac{11}{144}P)$. Setting a threshold at $5\sigma$ yields low probability for false positives, thus we can posit that for the correct $s_1$: 
$$\max(X_1,X_5,X_7,X_{11}) \geq \frac{7}{12}P+5\frac{\sqrt{11}\sqrt{7}}{12}\sqrt{P}$$
Since $P$ must be at least as large as this quantity, we can derive a minimum for $P$: $P \geq 77$, i.e. with $P=77$ true pairs (and no false pairs), the attack can succeed, and for $P>77$ it is guaranteed that if there are at least $\frac{7}{12}P+5\frac{\sqrt{11}\sqrt{7}}{12}\sqrt{P}$ true pairs, the correct $s_1$ value will survive the first phase. 

At the end of phase 1, the attacker obtains $s_1$. Additionally, since typically, only one of $X_1,X_5,X_7,X_{11}$ exceeds the threshold, say $X_I$ where $I \in \{1,5,7,11\}$, the attacker also learns $(\log_2 g)b \mod 12=I$, as well as $n_i$ for all pairs: 
$$n_i = $$
$$(\log_2(F'_i \oplus s_1)-\log_2(F_i \oplus s_1))((\log_2 g)b)^{-1} \mod 12$$
The attacker can also discard many false pairs -- pairs for which $n_i<2$ or $n_i>8$ are obviously false. Note that discarding a false pair does not disqualify the two individual outputs that form it. These values may still be useful for the second and third phases, which do not require pairs.

\textbf{Phase 2 -- Extracting $g$:}
We first note a property of Algorithm~\ref{alg:NetBSD-flowlabel}. The following two seeds yield the same output sequences, and are thus indistinguishable:
\begin{itemize}
    \item $(x,s_1,s_2,a,b,g)$
    \item $(-x \mod M,s_1,-M-s_2 \mod (N-1),a,-b \mod M,g^{-1} \mod N)$
\end{itemize}

To find the $g$ used in the PRNG, the attacker needs to enumerate over all possible $\varphi(N-1)=174744=2^{17.41}$ values of $g$. But practically, the attacker is not interested in the exact $g$ used in the PRNG. From the attacker's perspective, it is enough to be able to predict values, and for this, either the correct seed, or the negated seed suffice. In other words, the attacker can make do with enumerating over half of the $g$ space, i.e. $2^{16.41}$ values -- there is no need to cover both $g$ and $g^{-1}$. With $s_1$ from phase 1 and with the guess of $g$, the attacker obtains $x_i+s_2 \mod (N-1) = (\log_2 g)^{-1}\log_2(F_i \oplus s_1) \mod (N-1)$ for all $L$ values at the attacker's disposal (the attacker does not need pairs in this phase, just individual flow label values). 

Now the attacker can use the fact that $0 \leq x_i < M$, and that $M$ is significantly lower than $N-1$. This means that for the correct $g$, the set $\{(\log_2 g)^{-1}\log_2(F_i \oplus s_1) \mod (N-1)\}=\{x_i+s_2 \mod (N-1)\}$ will have a ``hole'' of size of at least $N-1-M$ in its coverage of $[0,N-1)$. A-priori, the attacker does not know where the hole is (because the attacker does not know $s_2$), but the {\em existence} of a hole whose size is at least $E=N-1-M=244332$ can be detected even without knowing $s_2$, by splitting the range of $N-1$ numbers into segments whose size is no larger than $\frac{E}{2}$. It is then guaranteed that one of the segments will be empty. In our case, the maximal segment size is $122166$, and the attacker needs 5 segments. For simplicity, the attacker can use equal-sized segments with $\lceil\frac{N-1}{5}\rceil=104854$ elements each. For each guess of $g$, the attacker goes over the the list of $(\log_2 g)^{-1}\log_2(F_i \oplus s_1) \mod (N-1)$ values, and each value is divided by the segment size to find the corresponding segment. The attacker stops if there are hits in all 5 segments. In such case, the attacker discards the guess of $g$ and moves on to the next one. For the correct guess of $g$, on the other hand, there will be at least one segment that is not a hit by all $L$ values. 

Interestingly, due to the algebraic properties of the series $(x_i)$, there may be incorrect $g'$ values that cause clustering in the values 
$$(\log_2 g')^{-1}\log_2(F_i \oplus s_1) \mod (N-1)=$$
$$(\log_2 g')^{-1}(\log_2 g)(x_i+s_2) \mod (N-1)$$
This can result in ``holes'' large enough to become false positives. When the $(x_i)$ series is truly random, this is statistically impossible, given large enough $L$ values (e.g. $L=100$), yet when $(x_i)$ is generated via a linear congruential generator, this does happen.

The following paragraph contains an in-depth explanation of this phenomenon, and the reader can skip it if these details are not necessary.

To demonstrate such false positive, define:
$$\gamma=(\log_2 g')^{-1}\log_2 g \mod (N-1)$$
Then the values that cluster are $(\gamma \cdot x_i \mod (N-1))$ (we can drop $s_2$ as it is merely a shift, and so does not affect clustering). Now consider a situation wherein $\gamma \cdot b$ and $\gamma \cdot M$ are ``small'' modulo $N-1$ (by that we mean they are close to either 0 or to $N-1$). Additionally, assume that 
$$\frac{M}{\gcd(a-1,M)}\big(\gamma \cdot \gcd(a-1,M) \mod (N-1)\big)$$
$$\ll (N-1)$$
For simplicity, let us assume that $n_i=1$ and that the series is extracted sequentially, i.e. $x_{i+1}=ax_i+b \mod M$. Now, $x_{i+1}-x_i \mod M = (a-1)x_i+b \mod M$ and $x_i(a-1) \mod M=x_i\frac{a-1}{\gcd(a-1,M)}\gcd(a-1,M) \mod M$. There are only $\frac{M}{\gcd(a-1,M)}$ possible values for this expression: $\{0,\gcd(a-1,M),2\cdot \gcd(a-1,M),\cdots,(\frac{M}{\gcd(a-1,M)}-1)\gcd(a-1,M)\}$. Define $\Delta_i=x_{i+1}-x_i$ (over the integers). Then $\Delta_i$ is either $t_i\cdot \gcd(a-1,M)+b$ or $t_i\cdot \gcd(a-1,M)+b-M$ or $t_i\cdot \gcd(a-1,M)+b-2M$, for some $0 \leq t_i < \frac{M}{\gcd(a-1,M)}$. Note that these expressions are over the integers (not modulo $M$). When we multiply this by $\gamma$ (modulo $N-1$), we get $\gamma \cdot \Delta_i \mod (N-1) = t_i\cdot (\gamma \cdot \gcd(a-1,M) \mod (N-1)) + (\gamma \cdot b \mod (N-1)) -u_i\cdot (\gamma \cdot M \mod (N-1))$ where $0 \leq t_i < \frac{M}{\gcd(a-1,M)}$ and $0 \leq u_i \leq 2$. Each $\Delta_i$ is therefore ``small'' (compared to $N-1$), so $x_i=x_0+\sum_{j=0}^{i-1} \Delta_j \mod (N-1)$ does not get too far from $x_0$.
This explains why there may be (few) false positives.

At the end of phase 2, the attacker has the correct $g$, perhaps along with some false positives. The attacker also knows which segment is inside the hole (the {\em empty segment}) for each $g$ candidate. There may be more than one such segment, and in such case the attacker can choose one segment arbitrarily as the empty segment.

\textbf{Phase 3 -- Eliminating the $g$ False Positives, and Finding $s_2$ (Partially):}
For each $g$ candidate, the attacker finds the minimum $S_{min}$ and maximum $S_{max}$ of the set $\{(\log_2 g)^{-1}\log_2(F_i \oplus s_1) \mod (N-1)\}=\{x_i+s_2 \mod (N-1)\}$ such that the range $[S_{min},S_{max}]$ form a continuous segment over the integers. If the set is entirely below the empty segment, then the minimum and maximum are taken from the set as usual. Likewise if the set is entirely above the empty segment. If there are elements in the set both above and below the empty segment, then the set starts above the empty segment, wraps around (courtesy of the modulo $(N-1)$) and ends below the empty segment. Therefore, the minimum is taken from the entries above the empty segment, and the maximum of the set is taken from the entries below the empty segment, to which we add $N-1$ in order to ``undo'' the wrap-around.

The attacker can easily eliminate almost all false positives for $g$ by discarding ranges larger than $M$, i.e. discard $g$ if $S_{max}-S_{min}+1>M$. Additionally, the attacker can apply a heuristic that eliminates additional, very rare false positives wherein the cluster is too ``tight'', i.e. form a range which is ``too small''. For example, the attacker can eliminate $g$ values that suggest $S_{max}-S_{min}+1<\frac{M}{2}$. Empirically, this leaves the attacker with a single (correct) $g$.

Now clearly, $s_2 \leq S_{min}$ and $S_{max} \leq s_2+M-1$. Put differently, $S_{max}-M+1 \leq s_2 \leq S_{min}$ (note that above, we discarded $g$ candidates for which $S_{max}-S_{min}+1>M$ i.e. $S_{max}-M+1>S_{min}$, therefore we know that $S_{max}-M+1 \leq S_{min}$; also, we allow negative values for $s_2$ here). We see that there are $M+S_{min}-S_{max}$ possible values for $s_2$. Since $S_{max}$ is distributed like a maximum of $L$ random variables uniformly distributed on $[0,M)$, $E(S_{max})=\frac{L}{L+1}M$, and similarly $E(S_{min})=\frac{1}{L+1}M$. Therefore, $E(M+S_{min}-S_{max})=\frac{2}{L+1}M$.

At the end of phase 3, the attacker has a correct $g$ and an average of $\frac{2}{L+1}M$ candidates for $s_2$, which is the range $S_{max}-M+1 \leq s_2 \leq S_{min}$. 

\textbf{Phase 4 -- Finding $a$ and Possible $b$ and $s_2$ Values:}
First, we make another observation on Algorithm~\ref{alg:NetBSD-flowlabel}. Concerning the intermediate values generated in line 13, it is easy to see that the output sequence induced by $(x,a,b)$ is at offset $+k$ (modulo $M$) from the output sequence induced by $((x-k) \mod M,a,(b+(a-1)k) \mod M)$ for any $k$. 
Let us assume that $s_2$ is correct, and so are $(x_0,a,b)$. Let us now look at an incorrect guess $s'_2=s_2+k \mod (N-1)$. This guess will lead to an (incorrect) internal state $x'_0=(x_0-k) \mod (N-1)$. If $\forall_{0\leq i <L} (0 \leq x_i-k < M)$, then $x'_i=(x_i-k) \mod M$, and we can pair $x'_0=x_0-k$ with $a'=a,b'=(b+(a-1)k) \mod M$ to form a sequence such that $\forall_{0\leq i <L}(x_i+s_2 \mod (N-1)=x'_i+s'_2 \mod (N-1))$. We see, therefore, that for every $k$ for which $\forall_{0\leq i <L} (0 \leq x_i-k < M)$, it is impossible to distinguish the correct sequence (with $s_2$) from a sequence generated with $s_2+k$. This gives rise, however, to an attack optimization. Instead of going over {\em all} $S_{max}-M \leq s_2 \leq S_{min}$, we can choose one arbitrarily, because every value in this range represents a possible $s'_2$ which keeps all $x'_i$ in $[0,M)$.

There is an important caveat here. While an arbitrary $S_{max}-M+1 \leq s'_2 \leq S_{min}$ is indistinguishable from the correct $s_2$ for all the flow label values already observed, it may happen that a future $x$ will fall outside the observed cyclic range $[S_{min}-s_2,S_{max}-s_2]$. In such case, an arbitrarily chosen $s'_2$ may fail to predict the correct PRNG output for $x$. One can think of this ``out of bounds'' $x$ value as redefining -- reducing -- the range of possible $s'_2$. At the extreme, if the attacker has outputs for $x_i=0,x_j=M-1$, then (and only then) the attacker will have a singleton ``range'' for $s'_2$, and thus $s'_2=s_2$.

The larger $L$ is, the less likely a random $x$ is to fall outside the range. In fact, the number of possible ``out of range'' values is identical to the size of the $s_2$ range, i.e. in expectancy of $\frac{2}{L+1}M$, therefore the probability for a random $x$ to fall ``out of range'' is $\approx \frac{2}{L+1}$. In fact, the attacker can choose $s_2=\frac{S_{max}-M+1+S_{min}}{2}$ to further reduce this probability. Note that since negative values of $s_2$ are possible if $S_{max}-M+1<0$, we may need to apply modulo $(N-1)$ to the chosen $s_2$.

After choosing $s_2$, the attacker uses pairs again, to find $a$ (the correct one) and $b$ (which matches the chosen $s_2$). Given $s_2, g, s_1$ and pairs $(F_i,F'_i)$ of consecutive flow label values, the attacker can calculate $x_i=(\log_2 g)^{-1}\log_2(F_i \oplus s_1)-s_2 \mod (N-1)$ for every $F_i$, and likewise $x'_i$ for $F'_i$. The attacker also knows $n_i$ for every such pair. The attacker starts with $a \mod 12=1$ and $b \mod 12=(\log_2 g)^{-1}I \mod 12$. At each step, the attacker extends $a$ and $b$ by $\log_2 3$ bits each. The attacker does this by defining $\tilde{a}=12m_a+(a \mod 12), \tilde{b}=12m_b+(b \mod 12)$ where $m_a,m_b \in \{0,1,2\}$ and going over all 9 combinations of $m_a,m_b$. For each combination of $m_a,m_b$, i.e. for each candidate of $a \mod 36=\tilde{a}, b \mod 36=\tilde{b}$ the attacker goes over all $P$ pairs with $n_i \ne 3 \land n_i \ne 6$  (it is easy to see that for pairs with $n_i=3$ or $n_i=6$, all such $\tilde{a}$ values yield correct results, hence they add no information), and counts how many pairs yield a consistent result, i.e.
$$\tilde{C}=$$
$$|\{i|x'_i \mod 36=\tilde{a}^{n_i}x_i+(\tilde{a}^{n_i-1}+\dots+1)\tilde{b} \mod 36\}|$$
Since the condition holds for all true pairs, if all pairs are true, we expect $\tilde{C}=\tilde{P}$ where $\tilde{P}=|\{i|n_i \ne 3 \land n_i \ne 6\}|$. If there are false pairs, we must use a statistical approach instead. For a false pair (and also, when the guess of $m_a,m_b$ is wrong), the probability for the condition to hold, for a single pair, is $\frac{1}{3}$. Hence, for a wrong guess, $\tilde{C}\sim\mathit{N}(\frac{\tilde{P}}{3}, \frac{\sqrt{2}}{3}\sqrt{\tilde{P}})$. We can use a threshold of $5\sigma$ again, and determine that a guess of $m_a,m_b$ is correct if
$$\tilde{C} \geq \frac{\tilde{P}}{3}+5\frac{\sqrt{2}}{3}\sqrt{\tilde{P}}$$
The attacker applies this technique repeatedly. First, the attacker calculates $a \mod 2^23^3$ and $b \mod 2^23^3$, then $a \mod 2^23^4$ and $b \mod 2^23^4$, etc., until the attacker has $a \mod 2^23^7, b \mod 2^23^7$. At this point, the attacker moves to applying the same technique with powers of two (this time, going over all pairs with $n_i \ne 4 \land n_i \ne 8$), to get $a \mod 2^33^7, b \mod 2^33^7$, etc., until the attacker gets $a \mod 2^73^7=a \mod M=a$, and $b \mod 2^73^7=b \mod M=b$. 

At the end of this phase, the attacker has $s_1,g,s_2,a,b$. With these values, and a recent sample of the NetBSD flow label, $F$, the attacker can predict the next value $F'$ of the flow label -- actually, the attacker can provide 7 possible values, one of them is guaranteed to be correct. For simplicity, we ignore the MSB prediction, which is trivial, and assume MSB=0. Then the attacker can calculate $x$ which was used for generating $F$: $x=(\log_2 g)^{-1}\log_2(F \oplus s_1)-s_2 \mod (N-1)$. The attacker can then enumerate over the 7 possible values of $n$, for each, the attacker calculates $x'=a^nx+(a^{n-1}+\dots+1)b \mod M$, and from that, $F'=s_1 \oplus (g^{x'+s_2} \mod N)$. In general, given $g,s_1,s_2,a,b$ and a sample $F$, the attacker can roll forward the PRNG indefinitely (up to reseeding, of course). 

\textbf{Covert Channel:}
This cryptanalytic result can be employed in order to implement a covert channel between a sender and a receiver. The receiver first establishes around 100 consecutive TCP connections with the target host, extracts the IPv6 flow labels, and obtains the next PRNG internal outputs using the cryptanalytic techniques. The sender then either establishes four TCP connections with the target host, to signal the bit ``1'', or does nothing, to signal the bit ``0''. Finally, the receiver establishes another TCP connection with the target host, and extracts the final IPv6 flow label. The receiver then matches this value to the series of eight possible future values (corresponding to the PRNG internal steps). If a match is found in positions 2-8 of the list, then the receiver infers that the sender did not establish four TCP connections with the target host (because each TCP connection incurs at least two internal steps, and the receiver's final connection also consumes at least two internal steps -- altogether 10 internal steps at the minimum). This is interpreted by the receiver as bit ``0''. Otherwise, the receiver assumes the bit sent was ``1''.

For host alias resolution use case, the attacker can improve the basic technique a bit, since the attacker can observe the flow label received from both endpoints, and match them to the same series (for endpoints that are mapped to the same host) or not (for endpoints that are mapped to different hosts).

\subsubsection{A Covert Channel via NetBSD TCP/IPv6 Flow Label Most Significant Bit}
\label{app:NetBSD-flowlabel-bit-flipping}
We now describe an attack against the NetBSD TCP/IPv6 flow label generation, which does not make any use of the specific algorithm that generates the least significant 19 bits of the flow label field. This is a more generic attack, compared to Appendix~\ref{app:NetBSD-flowlabel-cryptanalysis}, but at the price of being much less efficient, in terms of number of packets (TCP connections) required.

The attack is based on the observation that a ``passive'' side channel exists in the flow label generator, in the form of the timing of the MSB flip. When two TCP/IPv6 services are served from a single NetBSD host, the MSB in their flow label fields will flip simultaneously. But when the services are served from different NetBSD hosts, this is unlikely to happen.

We now calculate how many TCP connections are needed to force a reseed. In general, let $S$ be the internal steps the flow label PRNG executed while generating $n$ flow labels, then $S$ is a sum of $2n$ IIDs whose distribution is uniform on $\{1,2,3,4\}$. Therefore, $S\sim\mathit{N}(5n, \frac{\sqrt{10n}}{2})$, and $Prob(S \ge 200000)=\Phi(\frac{5n-200000}{\nicefrac{\sqrt{10n}}{2}})$. We can thus force an immediate algorithm reseed and MSB flip with probability $\approx 1$ by establishing $n=41000$ TCP connections with the target host.

This can be turned into an ``active'' covert channel. The sender and receiver agree on time $t_0$. Shortly before $t_0$, the sender establishes 41000 TCP connections with the target host and forces an algorithm reseed. Then every $\Delta t$ seconds the sender either establishes 41000 TCP connections with the target host to send the bit ``1'', or does nothing to send the bit ``0''. Every $\Delta t$ seconds the receiver establishes a single TCP connection with the target host, and observes whether the MSB bit flipped (signalling the bit ``1'') or not (``0''). $\Delta t$ should be chosen such that the sender can establish 41000 TCP connections with the target host during this time period. Special care must be taken if 180 seconds elapse since the last transmission of the bit ``1'', as this will cause the bit to flip due to a 180s timeout, and may thus result in a false positive (reading bit ``1'' by the receiver). To counteract this, the sender and receiver can decide to skip the time slice in which 180s elapsed since the previous bit ``1'' was transmitted.

\subsection{Linux NetFilter/nftables}
\label{app:nftables}
NetFilter is a network filtering framework which is built into the Linux kernel. NetFilter is controlled from user space by the nftables utility (the {\tt nft} command line). The NetFilter object hierarchy has tables that contain chains of rules.

Assuming a table named {\tt override-ipid} is created, e.g. by running:\\
\\
{\tt nft add table override-ipid}\\
\\
And assuming an output filter chain is created at priority ``raw'' (i.e. at the IPv4 level, but before fragmentation) is created, e.g. by running:\\
\\
{\tt nft \textquotesingle add chain override-ipid output \{type filter hook output priority raw;\}\textquotesingle }\\
\\
Then a safe configuration that only sets IPv4 ID=0 when DF=1 can be implemented by running:\\
\\
{\tt nft \textquotesingle add rule override-ipid output ip frag-off \& 0x4000 != 0 ip id set 0x0000\textquotesingle }\\
\balance
\\
A more aggressive configuration randomizes IP ID for ICMP packets, regardless of DF. This can be implemented by running the following (in addition to the above):\\
\\
{\tt nft \textquotesingle add rule override-ipid output ip protocol==1 ip id set numgen random mod 65536\textquotesingle }\\
\\
Finally, here is a very aggressive configuration, that unconditionally randomizes IPv4 ID:\\
\\
{\tt nft \textquotesingle add rule override-ipid output ip id set numgen random mod 65536\textquotesingle }\\
\\
Note that until recently, {\tt numgen random} was based on a weak Linux kernel random number generator (this implementation was attacked in \cite{Portland}). A more secure version is available in kernel versions 5.19-rc4 and above, and in LTS branches starting with versions 5.15.51 and 5.10.127. {\bf Using {\tt numgen random} in earlier versions is not recommended}.

\end{document}